\documentclass[groupedaddress,showpacs]{revtex4}
\usepackage{graphicx}
\pdfoutput=1 
\usepackage{amsmath}
\usepackage{color}
\usepackage{subfigure}
\usepackage{url}

\begin{document}

\title{Hydrodynamic Shock Wave Studies within a Kinetic Monte Carlo Approach}
\author{Irina Sagert$^1$, Wolfgang Bauer$^{2,3}$, Dirk Colbry$^3$, Jim Howell$^3$, Rodney Pickett$^3$, Alec Staber$^3$, Terrance Strother$^4$}
\affiliation{$^1$ Center for Exploration of Energy and Matter, Indiana University, Bloomington, Indiana, 47308, USA\\
$^2$Department of Physics and Astronomy, Michigan State University, East Lansing, Michigan, 48824, USA\\
$^3$Institute for Cyber-Enabled Research, Michigan State University East Lansing, Michigan 48824, USA\\
$^4$XTD-6, Los Alamos National Laboratory, Los Alamos, New Mexico 87545, USA}

\date{\today}
\begin{abstract}
We introduce a massively parallelized test-particle based kinetic Monte Carlo code that is capable of modeling the phase space evolution of an arbitrarily sized system that is free to move in and out of the continuum limit. Our code combines advantages of the DSMC and the Point of Closest Approach techniques for solving the collision integral. With that, it achieves high spatial accuracy in simulations of large particle systems while maintaining computational feasibility. Using particle mean free paths which are small with respect to the characteristic length scale of the simulated system, we reproduce hydrodynamic behavior. To demonstrate that our code can retrieve continuum solutions, we perform a test-suite of classic hydrodynamic shock problems consisting of the Sod, the Noh, and the Sedov tests. We find that the results of our simulations which apply millions of test-particles match the analytic solutions well. In addition, we take advantage of the ability of kinetic codes to describe matter out of the continuum regime when applying large particle mean free paths. With that, we study and compare the evolution of shock waves in the hydrodynamic limit and in a regime which is not reachable by hydrodynamic codes. 
\end{abstract}
%%%%%%%%%%%%%%%%%%%%%%%%%%%%%%%
%%%%%%%%%%%%%%%%%%%%%%%%%%%%%%%
%%%%%%%%%%%%%%%%%%%%%%%%%%%%%%%
\maketitle
\section{Introduction}
In fluid dynamics, when the characteristic length scale of a flow is smaller than the mean free path of its components, the continuum approximation which is assumed by the Navier-Stokes (NS) equations breaks down and the particle nature of matter must be taken into account \cite{Agarwal01}. Under these conditions, flows are said to be rarefied. The rarefaction of a flow is characterized by the Knudsen number $K$, which is defined as the ratio of the mean free path $\lambda$ to a characteristic length scale of the system $L$:
\begin{eqnarray}
K = \lambda / L .
\end{eqnarray}
While the continuum limit of hydrodynamics is applicable for $K \ll 1$, flows with $K > 0.01$ are not well described by the NS equations as these do not form a closed set in this regime \cite{Agarwal01}. Simulations of shock structures \cite{Agarwal01} and hypersonic flow \cite{Boyd95}, studies of space flight \cite{Li09} and nano-scale devices \cite{Yap12}, particle production in heavy-ion collisions \cite{Aichelin86,Bouras09, Bouras10, Bouras12,Kortemeyer95}, the dynamics of inertial confinement fusion (ICF) capsules \cite{Casanova91,Vidal93,Vidal95}, and neutrino-matter interactions in core-collapse supernovae (CCSN) \cite{Strother10} are all examples of research areas which require the ability to model flows with $K > 0.01$. In our work we are especially interested in the modeling of ICF capsules and CCSN.
%%%%%%%%%%%%%%%%%%%%%%%
%%%%%%%%%%%%%%%%%%%%%%%
\newline
Despite the enormous efforts poured into achieving thermonuclear ignition of ICF capsules at the National Ignition Facility (NIF), satisfactory yield has not yet been obtained. The ICF capsule experiments at NIF are guided by complex numerical simulations. Codes which suggest that the employed techniques should have been successful in igniting the nuclear fuel, all use continuum hydrodynamics to describe the entire capsule system. However, the so-called \textit{hot spot} of an ICF capsule where thermonuclear burn is expected to occur is approximately $25\:\mu$m across and has a density and temperature on the order of $100\:$g/cc and $10\:$keV, respectively \cite{Lindl04}. In this environment, the mean free path of a thermal deuteron, a critical component of the nuclear fuel, is on the order of a few micrometers and the continuum approximation is not applicable. Further investigations found that non-maxwellian ion distributions resulting from shock front kinetic effects that deplete the ion distribution of fast particles \cite{Casanova91,Vidal93,Vidal95,Larroche03,Bird94,Struckmeier95,Nordsieck67} may alter the hydrodynamics and thermonuclear yield of ICF capsules. Other studies suggest that capsule yield can be diminished by kinetic effects associated with strong self-generated electric fields which can broaden the shock fronts \cite{Amendt09,Amendt11}. With that, the physics of ICF capsules being possibly subject to strong non-equilibrium effects provides a strong motivation to develop a kinetic simulation code that is capable to fully resolve shock fronts in all regions of the capsules.
%%%%%%%%%%%%%%%%%%%%%%%
%%%%%%%%%%%%%%%%%%%%%%%
\newline
Many theories have been suggested to explain the explosion mechanism of CCSNe \cite{Colgate66,Bethe85,Burrows06,Bis71,LeBlanc70,Sagert09, Janka12}. The most commonly accepted model is based on energy deposition by neutrinos that are produced in the central high temperature and high density region of the collapsed stellar core into cooler less dense regions of infalling nuclear matter \cite{Colgate66,Bethe85, Janka07}. Simulations suggest that this so-called neutrino heating can power fluid instabilities that grow to large amplitudes and eventually power an explosion \cite{Blondin03,Herant94}. To study such a scenario, an accurate description of the supernova hydrodynamics and neutrino transport in three dimensions (3D) seems to be inevitable \cite{Fryer02,Nordhaus10,Takiwaki12,Hanke12}. While for 1D simulations, it is possible to solve the full Boltzmann equations for neutrino propagation \cite{Liebendoerfer05,Liebendoerfer04,Rampp02}, in two or three dimensions this approach becomes computationally too expensive and approximations have to be made \cite{Liebendoerfer09,Livne04,Marek09,Suwa09}. Monte Carlo neutrino transport has the potential to scale better for multi-dimensional calculations and has been suggested as an alternative approach for CCSNe studies (see \cite{Adbikamalov12} and references therein). Most simulations have been operating with nuclear equations of state which evolve only one representative heavy nucleus. More modern approaches apply equations of state that describe low density and low temperature nuclear matter as a full statistical ensemble of present nuclei ( see e.g. \cite{Hempel12, Buy13}). The inclusion and evolution of the latter could alter weak interaction processes and thereby the efficiency of neutrino heating as well as impact supernova nucleosynthesis \cite{Blinnikov11,Janka07,Sumiyoshi06,Langanke03}. It has been demonstrated in previous works that a kinetic approach has the capability to treat baryon and neutrino dynamics identically and thereby evolve an ensemble of nuclei as well as model out-of-equilibrium neutrino-matter interactions \cite{Strother10, Strother_thesis}. Preliminary studies conducted with such an approach found that out-of-equilibrium neutrino-matter interactions with nuclei close to the neutron drip line could significantly impact the dynamics of CCSN explosions \cite{Strother10, Strother_thesis}. However, more detailed calculations are necessary to confirm the observed effects.
%%%%%%%%%%%%%%%%%%%%%%%
%%%%%%%%%%%%%%%%%%%%%%%
\newline
It is the long-term goal of this work to contribute to the modeling of systems which are influenced by out-of-equilibrium phenomena, such as ICF capsules and CCSN. For this, we develop a kinetic code that is capable to resolve macroscopic 3D hydrodynamic flows and describe non-continuum behavior. Our approach is based on a combination of two kinetic methods for solving the Boltzmann equation - the Direct Simulation Monte Carlo technique (DSMC) and the Point of Closest Approach (PoCA) method. Together, these two methods can provide the computational feasibility to model large scale systems while maintaining spatial accuracy of the simulations. The latter is crucial for the study of shock fronts which play an important role for ICF capsules and CCSN dynamics. 
%%%%%%%%%%%%%%%%%%%%%%%
%%%%%%%%%%%%%%%%%%%%%%%
\newline
This work serves as an introduction to our kinetic Monte Carlo code as well as a demonstration of its ability to evolve shock fronts in 2D and 3D setups. After a brief introduction to kinetic methods in sections \ref{section:kinetic_methods}, we describe our simulation setup in section \ref{section:simulation_setup}. In section \ref{section:hydro_lim}, we present our results for shock wave modeling for matter in the hydrodynamic regime while section \ref{section:non_eq}  is devoted to shock fronts in matter with large particle mean free paths. Finally, we close with a summary and outlook in section \ref{section:conclusion}.
%%%%%%%%%%%%%%%%%%%%%%%%%%%%%
%%%%%%%%%%%%%%%%%%%%%%%%%%%%%
%%%%%%%%%%%%%%%%%%%%%%%%%%%%%
\section{Kinetic Methods}
\label{section:kinetic_methods}
\subsection{The Boltzmann equation}
\label{subsection:boltzmann_equation}
The governing equation for non-continuum flows is the Boltzmann equation of kinetic theory. It is a nonlinear integro-differential equation that describes the time evolution of a statistical distributions of particles in a fluid that undergo binary collisions. The Boltzmann equation is valid for all Knudsen numbers and leads to the the continuum description in the limit of small mean free paths \cite{Cercignani00}. For a fluid comprised of $N$ different particle species, the Boltzmann equation for the probability distribution function (PDF) of the $i$th particle species is \cite{Chapman70}:
%%%%%%%%%%%%%%%%%%%%%%%
\begin{eqnarray}
\frac{\partial f_i (\vec{r}, \vec{p}, t) }{\partial t} + \frac{ \vec{p}}{m_i} \cdot \nabla f_i (\vec{r}, \vec{p}, t) + \vec{F} \cdot \frac{\partial f_i (\vec{r}, \vec{p}, t)}{\partial \vec{p}} = I_{i,\mathrm{ coll}} .
\label{boltzmann} 
\end{eqnarray}
%%%%%%%%%%%%%%%%%%%%%%%
Hereby, $\vec{r}$, $\vec{p}$, and $m_i$ are the position, momentum, and mass of particle $i$ respectively. $\vec{F}$ is the external force field, and $f_i(\vec{r},\vec{p},t)$ is the number of particles of species $i$ that are found in a differential phase space neighborhood of $\vec{r}$ and $\vec{p}$ at time $t$. The term $I_{i,\mathrm{ coll}}$ on the right hand side of eq.(\ref{boltzmann}) takes into account the changes in $f_i$ that are induced by two-body collisions. $I_{i,\mathrm{coll}}$ is generally a sum of integrals involving the pre- and post-collision PDFs of all species present \cite{Chapman70}. It is this collision term that couples the $N$ differential equations and generally makes them very difficult to solve. The Boltzmann equation in not solvable by current analytical techniques. However, approximations can be employed to derive less complex equations. By making assumptions about the form of the PDFs, the Boltzmann equation can be used to derive approximate fluid dynamics equations that are of higher order in the Knudsen number than the NS equations \cite{Boyd95}. These higher-order fluid dynamic models are the so-called extended or generalized hydrodynamics equations. Another common approximate technique is to modify the Boltzmann equation itself. The most famous example of this is the Bhatnagar-Gross-Krook (BGK) \cite{Broadwell64} equation that replaces the collision term in eq.(\ref{boltzmann}) with a relaxation term:
%%%%%%%%%%%%%%%%%%%%%%%
\begin{eqnarray}
I_{i,\mathrm{ coll}}\rightarrow\frac{f_{M,i}-f_i}{\tau},
\label{relaxation}
\end{eqnarray}
%%%%%%%%%%%%%%%%%%%%%%%
where $f_{M,i}$ is the Maxwell distribution for the $i$th species, $f_i$ is the PDF of the $i$th species, and $\tau$ is a particle collision time. Despite the fact that the BKG equations are less complicated than the full Boltzmann equation, numerical techniques such as the lattice Boltzmann method or discrete velocity method \cite{Broadwell64,Li2004,Xu11,Liu12} are typically required to solve them.
%%%%%%%%%%%%%%%%%%%%%%%
%%%%%%%%%%%%%%%%%%%%%%%
\newline
In the absence of a general analytic solution technique for the full Boltzmann equation, simulation schemes that attempt to directly model the dynamics of a fluid using a large number of simulated particles have been developed. One of these approaches is the Molecular Dynamics (MD) method \cite{Alder59, ercolessi97} where each physical particle is represented by a simulated particle. Many advances have been made in MD calculations providing unique insight into fluid dynamics on molecular scales (see e.g. \cite{Graziani12,Kadau06,Murillo08,Schneider12}). However, the one-to-one physical-to-simulated particle ratio generally limits MD calculations to the numerical study of matter in small volumes.
%%%%%%%%%%%%%%%%%%%%%%%%%%%%%%%
%%%%%%%%%%%%%%%%%%%%%%%%%%%%%%%
%%%%%%%%%%%%%%%%%%%%%%%%%%%%%%%
\subsection{Test-Particle Method}
\label{subsection:tp-method}
In order to model the phase space evolution of systems that contain a large number of physical particles and therefore cannot make use of a MD-like simulation scheme, so-called test-particle methods are employed. These allow the physical-to-simulated particle ratio to greatly exceed one and thereby permit the modeling of macroscopic samples of fluids. Rather than attempting to discretize the entire phase space relevant to the simulated system and track the value of the phase space densities in each cell in time, test-particle methods only track the initially occupied phase space cells and represent them by test-particles. These are propagated in a way that models the physical evolution of the phase space. Hereby, the test-particle method approximates the phase space density with a sum over delta functions \cite{TobiasThesisReference54}
\begin{equation}
f(\vec{r},\vec{p},t)=\sum_{i=0}^{N} \delta^{3}\big(\vec{r}-\vec{r}_i(t)\big)\delta^{3}\big(\vec{p}-\vec{p}_i(t)\big) ,
\end{equation}
where $N$ is the total number of test-particles. The initial coordinates of the test-particles are determined by the initial conditions of the simulated system. Inserting the distribution function into the system's transport equations such as eq.(\ref{boltzmann}) generates a set of simple first-order linear differential equations: 
\begin{eqnarray} 
&& \frac{d}{dt} \vec{p}_i = \vec{F}(\vec{r}_i)+ \vec {\cal C}(\vec{p}_i), \: \: \: \frac{d}{dt} \vec{r}_i = \frac{\vec{p}_i}{m_{i}} 
\nonumber\\
&& i = 1,\ldots,N. 
\label{eom}
\end{eqnarray}
These govern the motion of the $i^{th}$ test-particle's centroid coordinates in the full six-dimensional phase space, whereas test-particles can interact with one another via one-body mean-field forces and scatter with realistic cross sections. Hereby, the mean-field force acting on the $i^{th}$ test-particle is expressed as $\vec{F}(\vec{r}_i)$ while $\vec{\mathcal{C}}(\vec{p}_i)$ symbolizes the effects that two-body collisions with other test-particles have on the $i^{th}$ test-particle's momentum. Similar to hydrodynamic simulations, scale and resolution concerns can occur when the spatial discretization of the simulated system is too crude. For example, failure to represent physical particles with a sufficient number of representative particles can prevent calculations from resolving details that are essential to the gross system behavior. For a system comprised of $N_{\mathrm{phys}}$ physical particles being modeled with $N$ test-particles, the ratio $N_{\mathrm{phys}}/N$ effectively determines a cutoff scale below which details cannot be resolved. The test-particle approach is applicable as long as $N$ is sufficiently large to capture the gross behavior of the system's phase space evolution. Convergence tests can ensure that small-scale phenomena which are impossible to resolve do not impact the phase space dynamics and/or can be taken into accout indirectly. 
%%%%%%%%%%%%%%%%%%%%%%%%%%%%%%
%%%%%%%%%%%%%%%%%%%%%%%%%%%%%%
%%%%%%%%%%%%%%%%%%%%%%%%%%%%%%
%%%%%%%%%%%%%%%%%%%%%%%%%%%%%%
%%%%%%%%%%%%%%%%%%%%%%%%%%%%%%
\subsection{Direct Simulation Monte Carlo Technique}
\label{subsection:dsmc}
The test-particle method typically employs probabilistic procedures to model two-body collisions and is therefore classified as a Direct Simulation Monte Carlo (DSMC) technique \cite{Bird63,Bird65,Bird94}. The primary approximation of the DSMC approach is a decoupling of the motion and the collisions of simulated particles during sufficiently small time intervals. DSMC techniques employ an operator splitting method to separate the Boltzmann equation into two processes: transport and collisions. The particle motion is calculated deterministically while two-body collisions are modeled probabilistically. Though variations exist between different DSMC implementations, a given iteration can be described by four basic processes: 1) sample flow field, 2) move the particles accordingly, 3) organize the particles into a scattering grid, and 4) model two-body collisions. The first two processes involve evaluating the flow field forces acting on the particles and then solving the resulting first-order linear differential equations of motion. In the third step the simulated volume is divided into a grid and particles are organized into the grid cells. Hereby, the grid cell size can depend strongly on the way the two-body collisions are modeled in the fourth step of the iteration.
%%%%%%%%%%%%%%%%%%%%%%%
%%%%%%%%%%%%%%%%%%%%%%%
\newline
A common approach to implement two-body particle scattering is to model collisions only between particles in the same scattering grid cell. Hereby, collision partners are selected randomly \cite{Bird94,Strother07} which leads to a high computational efficiency with a scaling of $N \log N$, whereas $N$ is again the total number of test-particles in the simulation. As demonstrated by Bird \cite{Bird70}, in the limits of infinite simulated particle count and vanishing scattering grid cell and time step size, DSMC algorithms that employ this type of scattering technique yield results that converge to known solutions of the Boltzmann equation. In particular, this scattering technique has been used extensively and with great success in the simulation of microscopic flows \cite{Bird70,Bauer05,Strother07} and the description of matter at low and high densities \cite{Alexander95}. However, to avoid collisions between particles that are too far apart from each other in a given time step, the size of the scattering cells has to be tightly connected to the mean free path of the physical particles. This can make DSMC scattering algorithms very expensive for flows with small $K$ and particle collision times much smaller the characteristic times of the simulated system. %%%%%%%%%%%%%%%%%%%%%%%%%%%%%%
%%%%%%%%%%%%%%%%%%%%%%%%%%%%%%
%%%%%%%%%%%%%%%%%%%%%%%%%%%%%%
%%%%%%%%%%%%%%%%%%%%%%%%%%%%%%
%%%%%%%%%%%%%%%%%%%%%%%%%%%%%%
%%%%%%%%%%%%%%%%%%%%%%%%%%%%%%
\subsection{Point of Closest Approach Method}
\label{subsection:poca}
In the Point of Closest Approach (PoCA) method, the collision between two test-particles is determined via their minimal distance $d_{\mathrm{min}}$. If the latter is reached during a time step $\Delta t$ the paths of two particles $A$ and $B$ intersect and a collision is possible. Typically, PoCA methods are computationally more intensive than DSMC calculations and scale as $N^2$. However, they enable a high spatial accuracy which is important for example in the numerical study of shock waves, and have been successfully used for of physical systems with limited number of test-particles, such as in the simulation of heavy-ion collisions \cite{Bertsch88,Cugnon81,Stoecker86}. To detect an intersection of particle paths, the relative position vector $\vec{r}_{\mathrm{rel}}$ at the current as well as the next timestep:
%%%%%%%%%%%%%%%%%%%%%%%
\begin{eqnarray}
\vec{r}_{\mathrm{rel}} (t)= \vec{r}_A (t) - \vec{r}_B (t) , \: \: \: \vec{r}_{rel} (t + \Delta t) = \vec{r}_A (t + \Delta t ) - \vec{r}_B (t + \Delta t)
\end{eqnarray}
%%%%%%%%%%%%%%%%%%%%%%%
has to be projected onto the corresponding relative velocity vector $\vec{v}_{\mathrm{rel}}$. This results in a crossing number:
%%%%%%%%%%%%%%%%%%%%%%%
\begin{eqnarray}
\chi = ( \vec{r}_{\mathrm{rel}} (t) \cdot \vec{v}_{\mathrm{rel}} (t) ) ( \vec{r}_{\mathrm{rel}} (t + \Delta t) \cdot \vec{v}_{\mathrm{rel}} (t + \Delta t) ) .
\label{c_value}
\end{eqnarray}
%%%%%%%%%%%%%%%%%%%%%%%
A negative value of $\chi$ indicates that within the time step $\Delta t$, the particles reach $d_{\mathrm{min}}$ and their paths cross. This is a necessary but not sufficient condition for a particle collision. The occurrence of the latter depends on the interaction probability which can be expressed via the interaction cross-section $\sigma$ or the particle mean free path $\lambda$. Both are tightly connected by the relation:
\begin{eqnarray}
\lambda = (\sigma n)^{-1} .
\end{eqnarray}
One possibility to implement the interaction probability into kinetic codes is by introducing an effective particle interaction radius $r_{\mathrm{eff}}$ which is related to $\lambda$ via: 
%%%%%%%%%%%%%%%%%%%%%%%
\begin{eqnarray}
\lambda = (4 \pi \: r_{\mathrm{eff}}^2 \: n)^{-1}. 
\label{reff_lambda}
\end{eqnarray}
%%%%%%%%%%%%%%%%%%%%%%%
Hereby, $n=N/V$ is the number of particles $N$ per volume $V$. Two particles $A$ and $B$ can only interact with each other if their distance at the point of closest approach is within the sum of the effective radii $d_{\mathrm{min}} \leq (  r_{\mathrm{eff}, A} + r_{\mathrm{eff}, B})$. A corresponding time for an overlap of the effective radii $t_o$ can then be calculated from: 
%%%%%%%%%%%%%%%%%%%%%%%
\begin{eqnarray}
\left| \vec{r}_{\mathrm{rel}} + \vec{v}_{\mathrm{rel}} \: t_o \right | &=& r_{\mathrm{eff}, A} + r_{\mathrm{eff}, B}, \\ 
| \vec{v}_{\mathrm{rel}} |^2 \: t_o^2 + 2 \: ( \vec{v}_{\mathrm{rel}} \cdot \vec{r}_{\mathrm{rel}} ) \: t_o + | \vec{r}_{\mathrm{rel}} |^2 &=& ( r_{\mathrm{eff}, A} + r_{\mathrm{eff}, B} )^2 
\label{collision_time}
\end{eqnarray} 
%%%%%%%%%%%%%%%%%%%%%%%
to: 
%%%%%%%%%%%%%%%%%%%%%%%
\begin{eqnarray}
t_{o \: 1,2} = \frac{1}{ |\vec{v}_{\mathrm{rel}} |^2} \left[ -  (\vec{v}_{\mathrm{rel}} \cdot \vec{r}_{\mathrm{rel}} )^2 \pm \sqrt{ ( \vec{v}_{\mathrm{rel}} \cdot \vec{r}_{\mathrm{rel}} )^2  - |\vec{v}_{\mathrm{rel}} |^2 \left( | \vec{r}_{\mathrm{rel}} |^2 - ( r_{\mathrm{eff}, A} + r_{\mathrm{eff}, B} )^2 \right) } \right] .
\label{collision_time2}
\end{eqnarray}
%%%%%%%%%%%%%%%%%%%%%%%
If the times $t_{o1}$ and $t_{o2}$ are real numbers a collision can take place. In case the particle distance is larger than the sum of the effective radii, that is $| \vec{r}_{\mathrm{rel}} | \gg  ( r_{\mathrm{eff}, A} + r_{\mathrm{eff}, B} )$, the particle cross-sections are too small and the collision times are imaginary. For systems in the hydrodynamic regime, the analysis of $t_o$ typically finds several potential collision partners for each particle. Once the latter have been determined, the final partners can be chosen either according to the shortest collision time or the smallest distance between scattering partners. In general, the potential impact that the screening of scattering partners by other test-particles has on physical scattering cross sections must be taken into account (see e.g. Kortemeyer et al. \cite{Kortemeyer96} and references therein).  For matter in the hydrodynamic limit, the scattering cross section is very large and the impact of screening thereby possibly negligible. However, possible effects can be avoided by requiring that the scattering partner which is selected for a given test-particle is always the closest of the potential candidates. Furthermore, by choosing final scattering partners according to the shortest distance, we can ensure a higher spatial resolution as well as minimize causality violation effects which can occur due to the finite distance of scattering partners when the latter are moving at relativistic speeds \cite{Kortemeyer95}. 
%%%%%%%%%%%%%%%%%%%%%%%
%%%                                                              %%%
%%%           SIMULATION SETUP            %%%
%%%                                                              %%%
%%%%%%%%%%%%%%%%%%%%%%%
\section{Simulation setup}
\label{section:simulation_setup}
\begin{figure}
\centering
\includegraphics[width=0.5\textwidth]{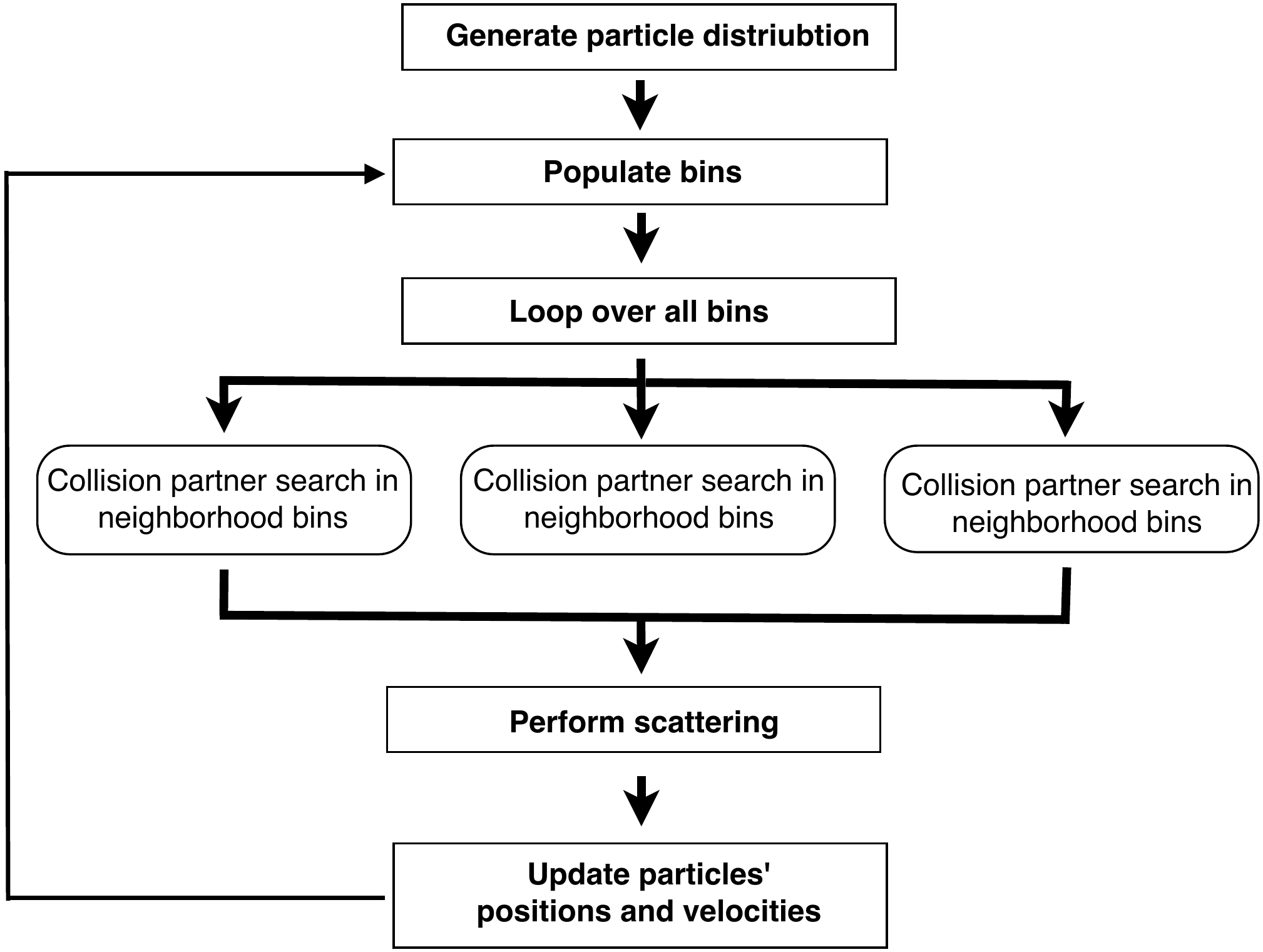}
\caption{Flow chart of particle simulation}
\label{flow_chart}
\end{figure}

A main motivation of our work is to set up an algorithm for collision detection which is efficient and scales well to simulations with a large test-particle number $N$. Hereby, the computationally most expensive step is the determination of interaction partners. A brute force approach would compare every particle to every other particle in the simulation. This is an $O(N^2)$ algorithm which could in theory run on $P$ processors for a running time that scales as $O(N^2/P)$. An alternative method is to assume that scattering partners are found via the PoCA method. A simple approach to the latter is to use a sorting algorithm which is $O( N\log N)$. Although there are methods for parallel sorting, this is still an expensive alternative and will not scale well to larger systems. For this work we choose to find scattering partners using spatial binning. Particles are placed into bins and interaction partners are searched via the PoCA method from particles of a particular bin and its neighbors. The binning component of the algorithm is linear with time and will scale as $O( N/P )$.  The search for scattering partners is now determined by looking only at other particles in the same bin or close neighbors. If the maximum number of particles in a bin $N_{\mathrm{bin}}$ is much smaller than the total particle number $(N_{\mathrm{bin}} < \log N < N)$ a significant decrease in the computation time of the algorithm can be expected. Furthermore, this approach scales linearly with $P$ for a total running time of $O( B \: N_{\mathrm{bin}}^2/P)$ whereas $B$ is the number of bins in the simulations. 
%%%%%%%%%%%%%%%%%%%%%%%
\begin{figure}
\centering
\includegraphics[width=0.7\textwidth]{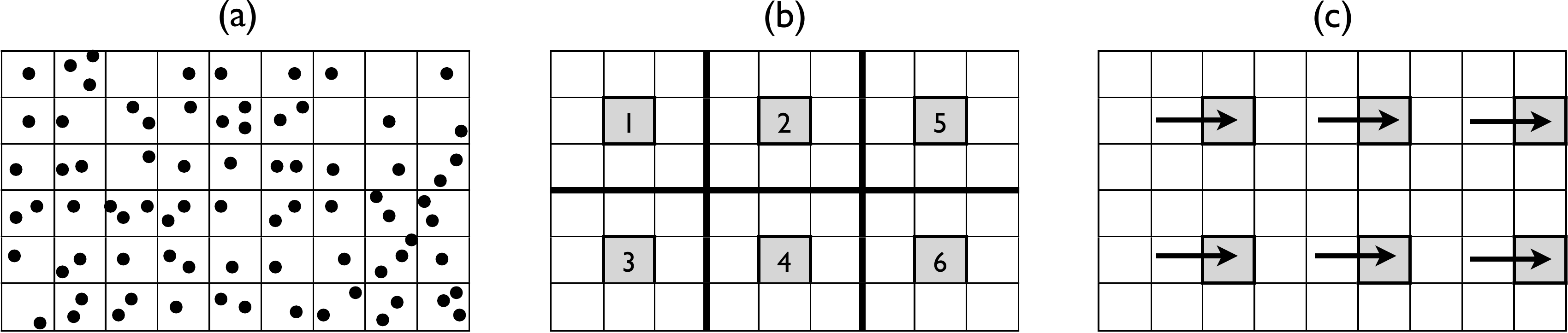}
\caption{(a) Visualization of a 2D simulation space with 54 bins filled with test-particles. (b) Parallel neighborhood binary collision detection is performed by six processors in the active bins (grey) and the surrounding 8-bin neighborhood. (c) Progress of the parallel scattering partner search via a simultaneous update of active bins for all processors.}
\label{bins}
\end{figure}
%%%%%%%%%%%%%%%%%%%%%%%
%%%%%%%%%%%%%%%%%%%%%%%
For this work the determination of scattering partners is parallelized using shared memory parallelization (OpenMP) whereas an approach which utilizes distributed memory parallelization (MPI) is currently in development. As is sketched in Figs.\ref{bins}, in the parallel setup, each processor is assigned a bin and its neighboring bins. For a 2D simulation, the neighborhood consists of 8 cells, while in a 3D setup the number of neighboring bins is 26. Bins cannot be shared among processors since two processors might assign the same collision partner to two different particles. The latter could lead to erroneous data in the update of particle's post-collision properties. Once the processors have completed the collision partner search they are assigned new bins with corresponding collision neighborhoods.
%%%%%%%%%%%%%%%%%%%%%%%
%%%%%%%%%%%%%%%%%%%%%%%
\newline
Figure \ref{flow_chart} gives a rough outline of the code structure. The simulation starts with the initialization of the particle distribution. Hereby, particles are assigned their properties such as spatial coordinates, velocities, and mass. 
%%%%%%%%%%%%%%%%%%%%%%%
%%%%%%%%%%%%%%%%%%%%%%%
The binning grid is set up with a specific size and number of bins in each dimension. To ensure that particles do not leave their scattering neighborhood within one timestep, we employ and adaptive timestep size $\Delta t (t)$ and couple it to the maximum particle velocity via:
\begin{eqnarray}
\Delta t (t) = \frac{ \Delta x} { v_{ \mathrm{max} } (t) }. 
\end{eqnarray}
Hereby, $v_{ \mathrm{max} } (t)$ is the speed of the fasted particle in the simulation at time $t$. This approach has a further advantage that time steps are not unnecessarily small as long as the difference between $v_{ \mathrm{max} } (t)$ and the average particle velocity is not too large. At the beginning of each time step, the particles are sorted into bins. Within a loop over all bins, the collision partner search is performed in parallel with the following criteria for scattering detection: 
%%%%%%%%%%%%%%%%%%%%%%%
\begin{enumerate}
\item Negative crossing number $\chi$ during $\Delta t$
\item Real times for the overlap of effective radii $t_{o, \: 1}$ and $t_{o, \: 2}$ for a given $\lambda$
\item Minimum distance between final collision partners 
\end{enumerate}
%%%%%%%%%%%%%%%%%%%%%%%
Once collision partners are assigned to each other, we perform the scattering in the center of mass frame of each colliding pair. This is currently done in a Newtonian formalism but can be extended to a relativistic regime \cite{Sorensen94,Kortemeyer95, Bouras09, Bouras10, Bouras12}. As is usually done in DSMC methods, rather than determining the exact outgoing velocity vectors under e.g. the assumption of spherical test-particles, we orient the outgoing post-collision velocity vectors randomly and opposite to each other in space. Once scattering has been performed for all colliding particles, we update the particle positions. For that, we determine the time at which the interacting particles reach their point of closest approach:
\begin{eqnarray}
t_{ \mathrm{poca} } = - \frac{ \vec{r}_{ \mathrm{rel} } \cdot \vec{v}_{ \mathrm{rel} } } { \left| \vec{v}_{ \mathrm{rel} } \right|^2 }.
\end{eqnarray}
Particle positions are then updated via:
\begin{eqnarray}
\vec{r} (t + \Delta t) = \vec{r} (t) + \vec{v}_{\mathrm{old}} \: t_{\mathrm{poca}} + \vec{v}_{\mathrm{new}} ( \Delta t - t_{\mathrm{poca}}) ,
\end{eqnarray}
where $\vec{v}_{\mathrm{old}}$ is the pre-collision velocity. For non-interacting particles, the new positions are determined simply via:
\begin{eqnarray}
\vec{r }(t + \Delta t) = \vec{r}(t) + \vec{v}_{\mathrm{old}} \: \Delta t. 
\end{eqnarray}
For large values of $N_\mathrm{bin}$ it is not unlikely that two particles find the same collision partner \cite{Kortemeyer95}. In our current approach we chose to disregard such configurations and only to consider unambiguously determined interaction partners. This can lead to a substantial reduction of possible collisions, especially in the continuum regime \cite{Kortemeyer95}. However, we find that for small particle mean free paths, typically 50\% or particles interact with each other during each time step $\Delta t$. This is a reasonable value to ensure prompt equilibration of the system as will be shown in the next section where we apply the kinetic particle code to hydrodynamic shock wave studies. 
%%%%%%%%%%%%%%%%%%%%%%%%%%%%
%%%%%%%%%%%%%%%%%%%%%%%%%%%%
%%%%%%%%%%%%%%%%%%%%%%%%%%%%
\section{Numerical studies in the hydrodynamic limit}
\label{section:hydro_lim}
As mentioned in the introduction, an advantage of kinetic approaches is their ability to capture the behavior of matter for all Knudsen numbers. In this work we want to test the capability of our scattering partner search algorithm to reproduce hydrodynamic behavior especially with regard to the evolution of shock waves. The latter are often used as test cases to study the capability of hydrodynamic codes to capture steep gradients and discontinuities. Furthermore, many test problems have analytic solutions which allow a qualitative comparison of different codes. The test suite which we present in this work consists of the Sod  \cite{Sod78}, the Noh \cite{Noh87}, and the Sedov tests \cite{Sedov59}. All tests were performed as 2D and 3D simulations. The Sod shock test is a classical Riemann problem which describes a propagating shock front. It is considered to be an elementary benchmark test that all codes designed to model shock waves should pass. The setup of the Noh test is a gas that is collapsing towards the origin which leads to the development of a standing shock front that continuously grows in size. This problem tests a code's ability to convert kinetic into internal energy and reveals anomalous wall heating which is present for some codes that implement artificial viscosity. Similarities between the infall stage of a core-collapse supernova and the setup of the Noh test, make the latter an interesting test case. The Sedov problem requires a code to propagate strong shocks in non-planar geometry. It creates and evolves a spherically symmetric blast wave after an energy deposition in the center of the simulation space. Also here, strong parallels to CCSNe are present since the Sedov shock front is very similar to the one that is created in a type-II supernova after the so-called bounce of the in-falling iron core.      
%%%%%%%%%%%%%%%%%%%%%%%
%%%%%%%%%%%%%%%%%%%%%%%
\newline
Our tests are carried out with $O(10^7)$ test-particles placed in a simulation box which is divided into $(10^4 - 10^6)$ equally sized bins. Hereby, we distinguish between two types of grids. The first grid is used for the collision partner search and is divided into \textit{calculation} bins ($c$-bins). Generally, a large number of calculation bins and thereby small number of test-particles per bin is preferred as it decreases the computational time. However, since the width of a $c$-bin $\Delta x$ is also the maximal distance that a test-particles can travel in one time step, an increase in the number of $c$-bins, decreases $\Delta x$ resulting in a larger number of time steps which are required for the simulation to reach a specific time. For the analysis of our results we use \textit{output} bins ($o$-bins). Their number can be different from the number of calculation bins and is typically chosen to be small since a large number of particles per output bin reduces fluctuations in the calculated thermodynamic quantities. For systems in the continuum regime we set the particle mean free path to be a small fraction of the $c$-bin size $\lambda = 10^{-3} \Delta x$ to ensure small Knudsen numbers. Since the aim of this study is to present a proof-of-principle that our kinetic code is capable to reproduce shock wave dynamics in a multi-dimensional setup we do not choose a specific equation of state or particle mean-field potential but model the particle interactions via elastic scattering. Therefore, matter behaves as an ideal gas with the number of degrees of freedom $f$ given by the number of dimensions in the simulation. With that, the heat capacity ratio $\gamma = 1 + 1/f$ becomes $\gamma = 2$ for 2D simulations and $\gamma = 5/3$ in the case of three dimensions. For future studies we will employ more flexible heat-capacity ratios (for an approach to implement arbitrary values of $\gamma$ see e.g. \cite{Isaka06}) and particle mean-field potentials as it is widely applied in particle codes. 
%%%%%%%%%%%%%%%%%%%%%%%
%%%%%%%%%%%%%%%%%%%%%%%
\newline
The thermodynamic quantities which are calculated and compared to analytic solutions are the particle number density $n$, the pressure $p$, the velocity $v$, and the temperature $T$.  The density is simply given by the number of particles per corresponding volume $n = N/V$. The velocity of matter is determined either as bulk velocity:
\begin{eqnarray}
v_b &=& \frac{1}{N} \sqrt{ v_{b,x}^2 +  v_{b,y}^2 + v_{b,z}^2 }, \: \: \: v_{b,\alpha} = \sum\limits_{i=1}^N v_{i,\alpha}, \: \: \: \alpha=x,y,z .
\label{vel_b}
\end{eqnarray}
or radial velocity: 
\begin{eqnarray}
v_r =  \frac{1}{N} \sum\limits_{i=1}^N \frac{\vec{v}_i \cdot \vec{r}_i }{ \left| \vec{r}_i \right| } 
\label{vel_r}
\end{eqnarray}
depending on the geometry of the simulation. The pressure is derived from the stress tensor of dense gases \cite{Mulero08,Irving50}:
\begin{eqnarray}
\mathbf{P}_{\alpha \beta} = - \left( \sum_i m \left( v_{i,\alpha} - v_{b,\alpha} \right) \left( v_{i,\beta} - v_{b,\beta} \right)  + \frac{1}{2} \frac{1}{\Delta t} \sum_i \sum_{i\neq j} r_{ij, \alpha} \: \Delta p_{i, \beta} \right).
\label{stress}
\end{eqnarray}
Hereby, $i$ and $j$ are interacting particles and $\vec{r}_{ij} = \vec{r}_i - \vec{r}_j$ is their distance at the point of closest approach. The momentum which is transfered during the interaction is given by $\Delta \vec{p}_i = m \left( \vec{v}_{i,\mathrm{new}} -\vec{v}_{i,\mathrm{old}}\right)$. The stress tensor $\mathbf{P}_{\alpha \beta}$ is composed of a kinetic and a potential contribution whereas the first term in eq.(\ref{stress}) is the kinetic part. It dominates for gases and is induced by the momenta carried by the particles around their bulk motion. It is also the contribution of the stress tensor which results in the usual thermal pressure. The second term is the potential part which arises due to momentum transfer from particle interaction and dominates the pressure in liquids. Applying eq.(\ref{stress}) to particles in a volume $V$ results in a corresponding pressure of: 
\begin{eqnarray}
p_{2D} = - \frac{1}{V} \frac{\mathbf{P}_{xx} + \mathbf{P}_{yy}}{2}, \: \: \: p_{3D} = - \frac{1}{V} \frac{\mathbf{P}_{xx} + \mathbf{P}_{yy} + \mathbf{P}_{zz}}{3}. 
\end{eqnarray}
Similar to the kinetic pressure contribution, we determine the temperature via the root-mean-square velocity:
\begin{eqnarray}
v_{\mathrm{rms}} =  \frac{1}{N} \left( \sum\limits_{i=1}^N \left[ (v_{i,x} - v_{b,x})^2  + (v_{i,y} - v_{b,y})^2 + (v_{i,z} - v_{b,z})^2 \right] \right)^{1/2},  
\end{eqnarray}
whereas $T = v_{\mathrm{rms}}^2 m/f$. To analyze the simulation results we determine $n$, $p$, $v$, and $T$ per bin and as averages for a given distance from the origin. Our simulations are performed on the High Performance Computer Center at Michigan State University. Depending on the setup and the total number of particles we either apply 8 CPUs (2.4 GHz Intel Xeon E5620) or 30 CPUs (2.66 GHz Intel Xeon E8837) for shared memory parallelization. In the following we will describe the outcome of the performed tests as well as the corresponding running times, i.e. wall-times. It should be noted that the latter and the CPU scaling can strongly depend on the distribution of test-particles i.e. load of processors. Ideally the latter should be homogeneous with a low number of particles per bin. For systems which involve a large number of particles concentrated in only a few bins the computational load can be unbalanced which would results in a decrease of speed-up and computational efficiency. Possible solutions to ensure a more homogeneous load include the usage of adaptive grids or the dynamic update of CPUs in the parallel scattering partner search. However, even with a fixed homogeneous number of particles per bin, the running time of a simulation will increase with higher number of bins if the number of CPUs is not changed accordingly. We therefore expect higher computational times for simulations which contain particle accumulations in comparison to tests with homogeneous particle distributions as well as when going from 2D simulations to 3D setups since the latter generally require a larger number of bins. 
%%%%%%%%%%%%%%%%%%%%%%%
%%%%%%%%               %%%%%%%%%%%
%%%%%%%%   CPU   %%%%%%%%%%%
%%%%%%%%               %%%%%%%%%%%
%%%%%%%%%%%%%%%%%%%%%%%
\subsection{Equilibration and Speed-up}
\label{maxwell_boltzmann}
To study the equilibration rate in dependence of the test-particle mean free path, we set up a two-dimensional simulation space with $N=2.0 \times 10^7$ test-particles distributed over $1000 \times 1000$ $c$-bins with $0 \leq x,y \leq 4$. All particles are initialized with the same absolute velocity $v_{\mathrm{in}}=10$ and random velocity vector orientation. Varying the mean free path between $\lambda = 10^{-3} \: \Delta x$, $1.0 \: \Delta x$, and $10 \: \Delta x$, we determine the particle velocity distribution over the first 400 time steps whereas the time step size is chosen to be constant with $\Delta t = \Delta x / 60$. For systems in equilibrium the particle velocity distribution should promptly reach the Maxwell-Boltzmann form.
%%%%%%%%%%%%%%%%%
\begin{figure}
\centering
\subfigure{
\includegraphics[width=0.4\textwidth]{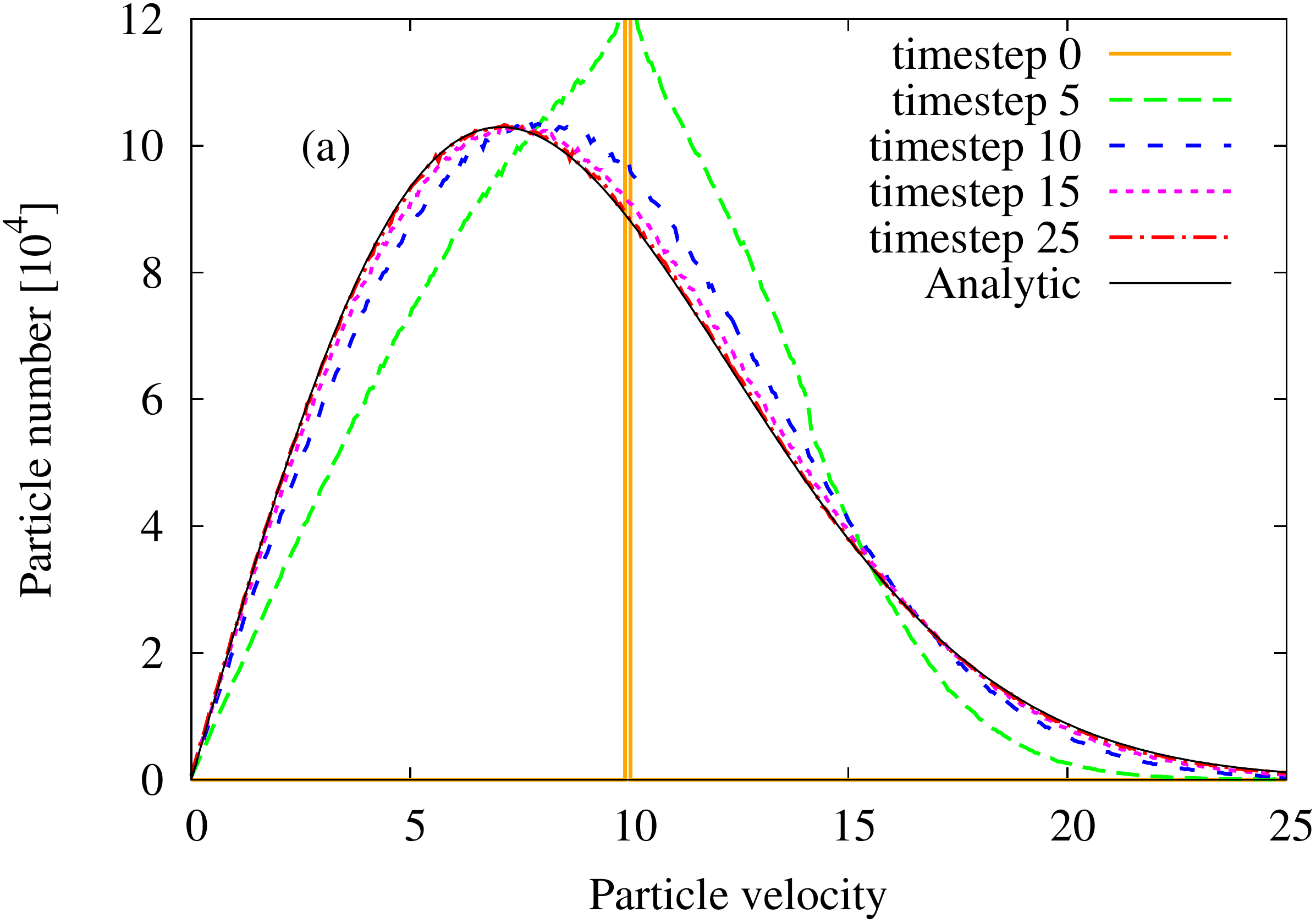}}
\subfigure{
\includegraphics[width=0.4\textwidth]{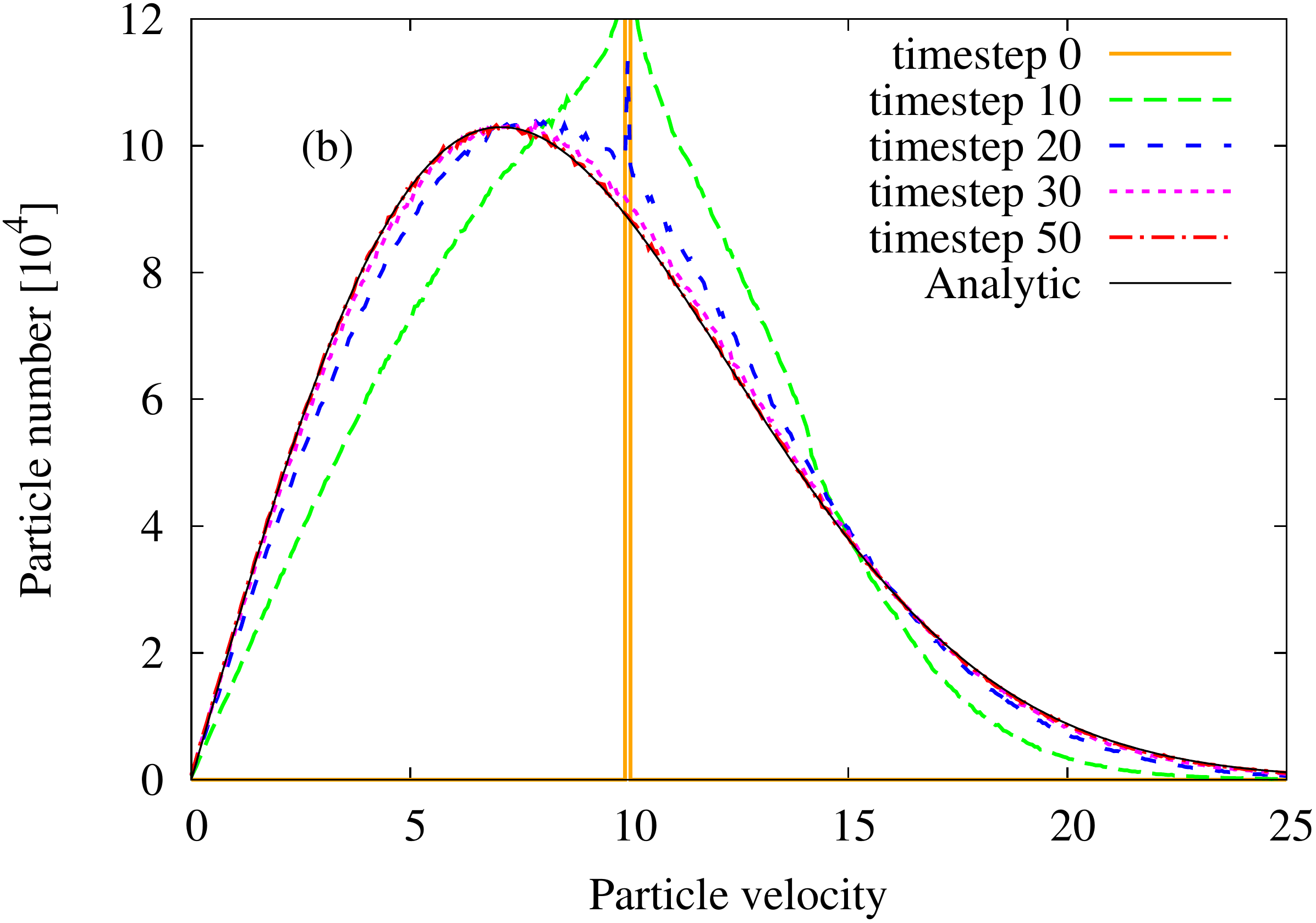}}
\subfigure{
\includegraphics[width=0.4\textwidth]{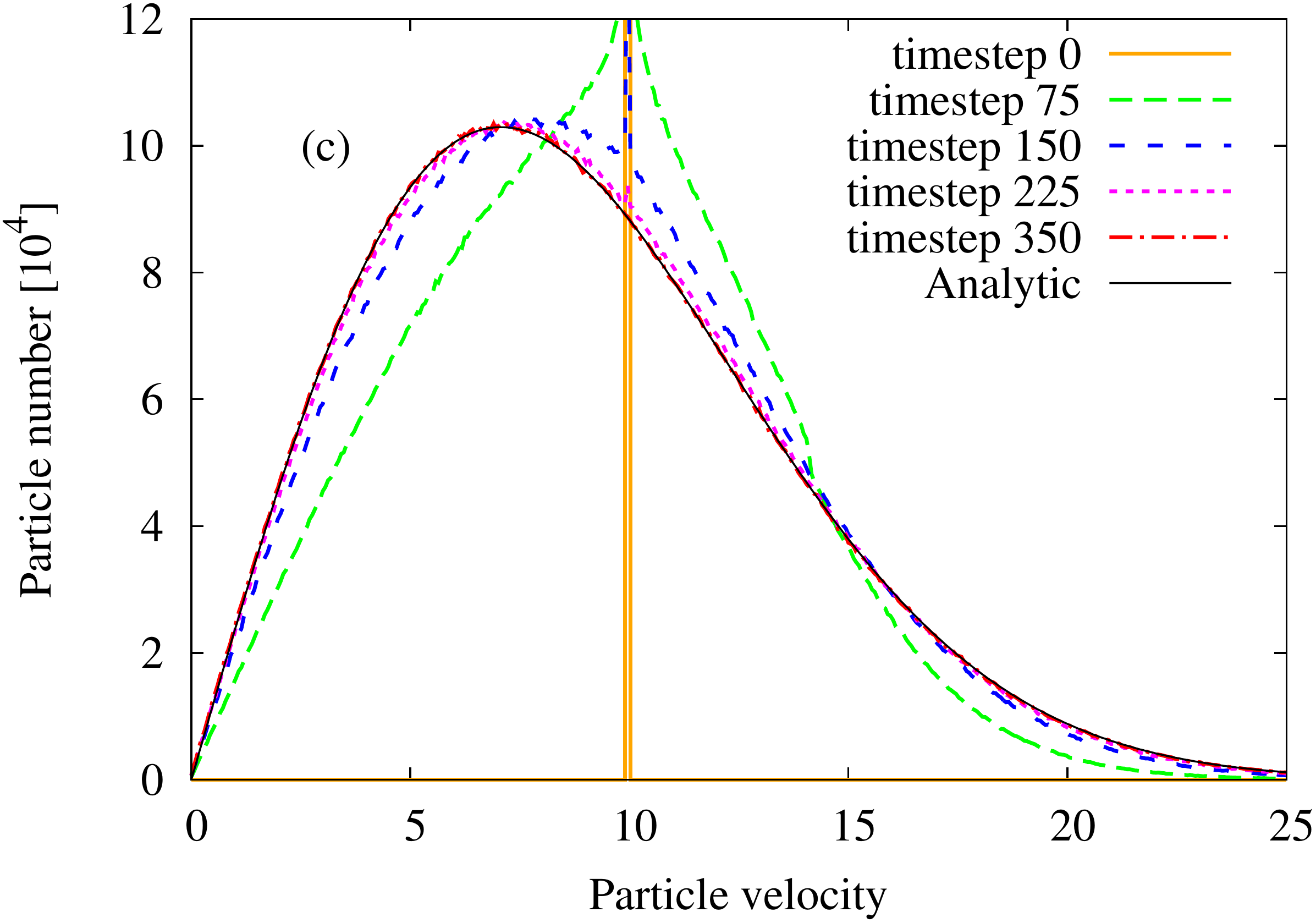}}
\subfigure{
\includegraphics[width=0.4\textwidth]{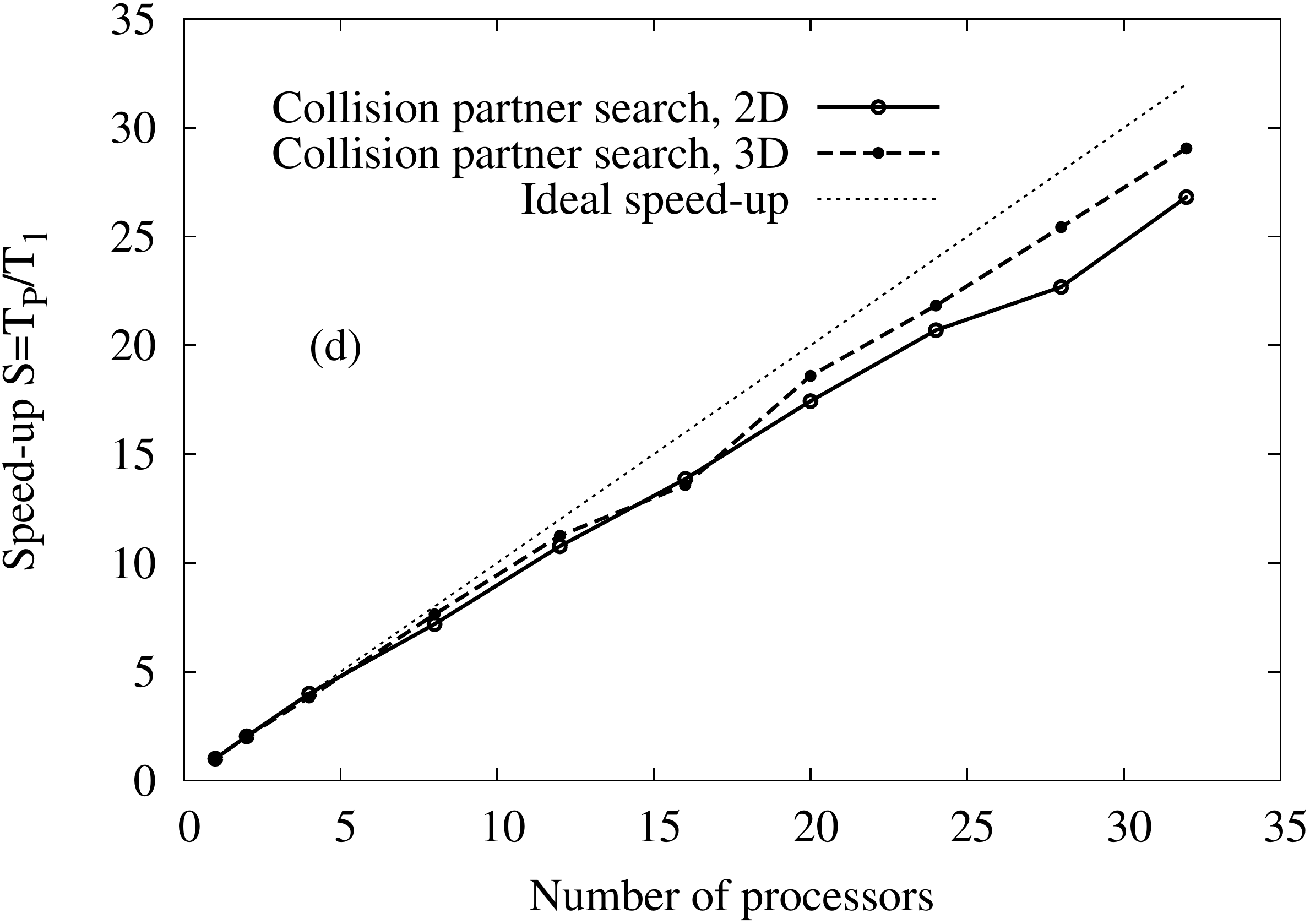}}
\caption{Velocity distributions with $N = 2.0 \times 10^7$, $1000 \times 1000$ bins. From left top to right bottom: (a) $\lambda = 10^{-3} \: \Delta x$, (b) $\lambda = 1.0 \: \Delta x$, and (c) $\lambda = 10 \: \Delta x$ for different simulation times together with the analytically obtained Maxwell-Boltzmann velocity distribution. (d) Speed-up of the 2D collision partner search for the first 100 time steps as a function of the number of CPUs with $\lambda = 10^{-2} \: \Delta x$. In addition we plot the speed-up for a 3D calculation for the first 20 time steps applying the same initialization as the 2D simulation but with $100 \times 100 \times 100$ bins.}
\label{maxwell_boltzmann_figures}
\end{figure}
Figures \ref{maxwell_boltzmann_figures}(a)-(c) show the particle number as a function of absolute velocity for different values of $\lambda$ together with the analytic prediction. It can be seen that the simulation with $\lambda = 10^{-2} \: \Delta x$ reaches the Maxwell-Boltzmann velocity distribution within the first $15-25$ time steps. As previously mentioned, for simulations with small particle mean free paths, i.e. large collision rates, many interactions are not performed as they are not unambiguously determined between two collision partners but can involve more particles \cite{Kortemeyer95}. However, despite this exclusion of interactions, we find that for $\lambda = 10^{-3} \: \Delta x$ generally about 50\% of the particles scatter with each other per time step $\Delta x$. For the simulation applying $\lambda \gtrsim 1.0 \: \Delta x$ the number of interacting particles per time step reduces to about 25\%. Consequently, as can be seen in Fig. \ref{maxwell_boltzmann_figures}(b), the simulation requires about twice as many time step to reach the equilibrium configuration. However, a particle mean free path of $\lambda = 1.0 \: \Delta$ is still within the width of the collision neighborhood. Therefore, despite a reduced particle collision rate, a mean free path of $\lambda = 1.0$ can still be expected to lead relatively promptly to the continuum regime. For $\lambda = 10 \: \Delta x$ equilibration is reached only after $t \sim 350 \: \Delta t$ as only about 3\% of the particles undergo collisions within one time step. With that, we will use small values of $\lambda = 10^{-3} \: \Delta x$ in the upcoming shock wave simulations whereas $\lambda = 1.0 \: \Delta x$ and $\lambda = 10 \: \Delta x$ will be applied as benchmark values in our shock studies for matter with large $K$.  
%%%%%%%%%%%%%%%%%%%%%%%
%%%%%%%%%%%%%%%%%%%%%%%
\newline
As previously discussed, our scattering partner search algorithm is designed to work in parallel. To test the speed-up of the code we perform timing tests for the above velocity equilibration study for the first 100 time steps using $\lambda = 10^{-3} \: \Delta x$. The computational speed-up is defined as $S=T_{P}/T_1$, where $T_{P}$ is the wall-clock performance time of an algorithm with $P$ number of processors, while $T_1$ is the time when only one processor is used. The results for the first 100 time steps of above 2D equilibration test using $P = 1 - 32$ are shown in Fig.~\ref{maxwell_boltzmann_figures}(d) in comparison to an ideal speed-up where $S = P$. The wall-times for the Maxwell-Boltzmann equilibration with 8 processors range from 0.6-0.8 hours depending on the particle mean-free-path and decreasing for larger values of $\lambda$. The latter effect is due to the smaller particle effective radii (see eq.(\ref{reff_lambda}) ). A decrease in $r_{\mathrm{eff}}$ leads to higher exclusion of particles as possible collision partners in step 2 of the collision detection (see section\ref{section:simulation_setup}) and thereby a reduction of time that the code spends in the collision partner search. Though the performed simulations show a deviation from linear speed-up with increasing number of processors the overall behavior is close to ideal and therefore promising for implementations of our algorithm on larger systems with thousands of cores. We also perform a timing study in 3D with a similar setup as in the 2D equilibration study, i.e. $N=2.0 \times 10^7$, $v_{\mathrm{in}}=10$, $\lambda = 10^{-2} \: \Delta x$, and $0 \leq x,y,z \leq 4.0$. However, now with $100 \times 100 \times 100$ $c$-bins and only over the first 20 time steps. As can be seen in Fig.~\ref{maxwell_boltzmann_figures}(d), the 3D simulations shows a similar scaling of the computational time with $P$, being close to an ideal behavior but with growing deviations for larger number of processors. We want to point out that the overhead time, e.g. for the sorting of test-particles into the bins, the performance of collisions, the update of post-collision particle properties, and the output of particle data, is not included in the timing study. For small systems, where the simulation spends an almost equal amount of time in different parts of the code, the deviation of the speed-up from ideal behavior can be significant. However, our focus lies on large systems where the overall computational time is dominated by the collision partner search which therefore determines the scaling behavior of the code.
%%%%%%%%%%%%%%%%%%%%%%%
%%%%%%%%               %%%%%%%%%%%
%%%%%%%%   SOD   %%%%%%%%%%%
%%%%%%%%               %%%%%%%%%%%
%%%%%%%%%%%%%%%%%%%%%%%
\subsection{Sod test}
\label{sod_test}
Our series of shock wave studies starts with the Sod shock tube test \cite{Sod78}. This test is a classical a Riemann problem and has been performed by various hydrodynamic and kinetic algorithms \cite{Tasker08, Fryxell00,Crouseilles04,Li04,Smith09,Gan08,Xu10,Degond10,Dimarco13}. The simulation space is split into two equally sized domains with the number density, pressure, and bulk velocity ratios of the right and left-hand sides given by:
\begin{eqnarray}
n_r = 0.125 \: n_l,  \: \: \; p_r = 0.1 \: p_l, \: \: \: v_{b,r} = v_{b,l}  = 0.
\label{np_sod}
\end{eqnarray}
At the beginning of the simulation, matter from the high density region starts to propagate into the left domain leading to the development of a shock front. The latter is followed by a contact discontinuity and an expansion wave. Since the contact discontinuity separates fluids with different entropies, it manifests itself in the density and temperature while the pressure and velocity stay constant across the separation. We perform a two-dimensional and a three-dimensional simulation, whereas for both tests, we chose reflective boundary conditions and initialize the particle velocities via Maxwell-Boltzmann distributions and random orientations. Results of the shock front evolution are compared to analytic solutions obtained by the exact\_riemann.f code \cite{Fryxell_riemann}.
%%%%%%%%%%%%%%%%%%%%%%%
%%%%%%%%%%%%%%%%%%%%%%%
\newline
The 2D simulation is carried out with $N = 2.0 \times 10^7$ test-particles in a simulation space with $0 \leq x\leq 7$ and $0 \leq y\leq 1.75$, $2000 \times 500$ $c$-bins, and $400 \times 100$ $o$-bins. The particle mean free path is set to $\lambda = 10^{-3} \: \Delta x$, whereas $\Delta x = 7.0/2000 =3.5 \times 10^{-3}$. 
%%%%%%%%%%%%%%%%%%%%%%%
\begin{figure}
\centering
\includegraphics[width=0.90\textwidth]{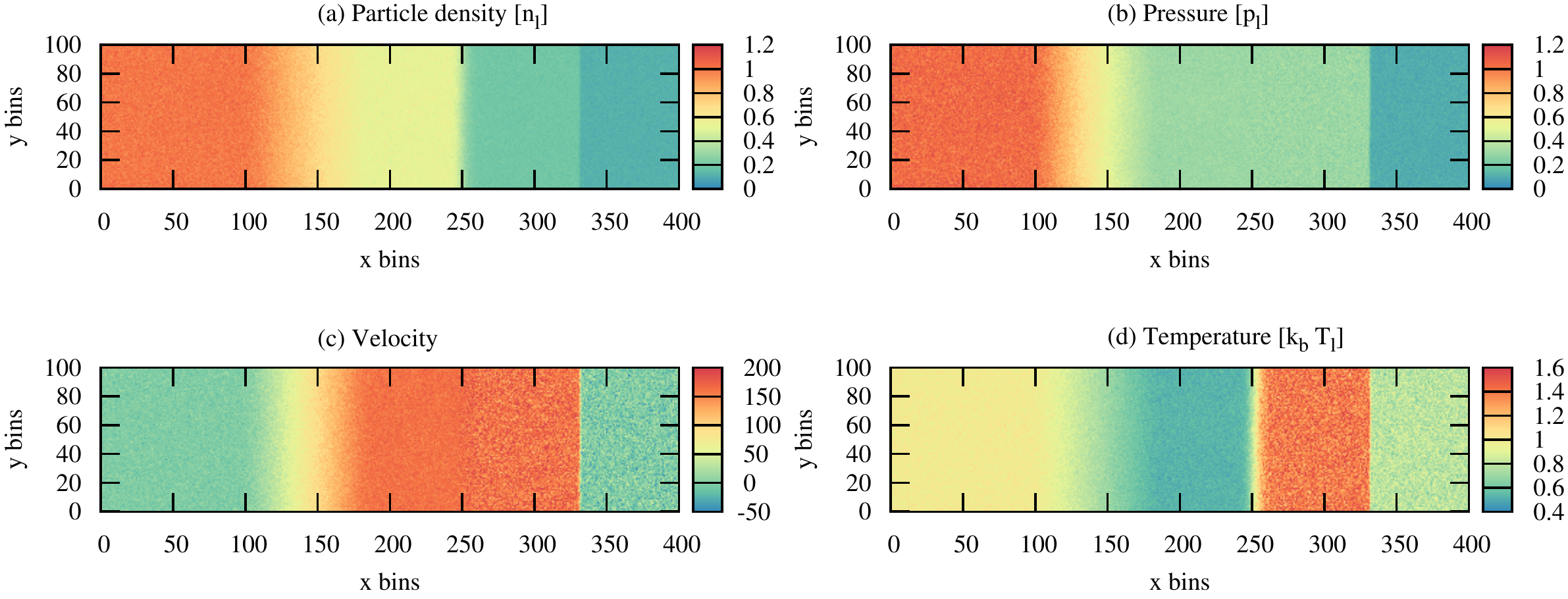}
\caption{2D Sod test: $N = 2 \times 10^7$ test-particles distributed over $400\times 100$ $o$-bins with $\lambda = 10^{-3} \: \Delta x$. (a) Normalized particle density, (b) normalized pressure, (c) bulk velocity, and (d) normalized temperature per o-bin at time $t \sim 5.6 \times 10^{-3}$. Normalizations are done with $n_l \sim 2.906 \times 10^{6}$, $p_l \sim 1.305 \times 10^{11}$, and $T_l = 4.5 \times 10^4$.}
\label{sod_2D}
\vspace{1cm}
\centering
\includegraphics[width=0.8\textwidth]{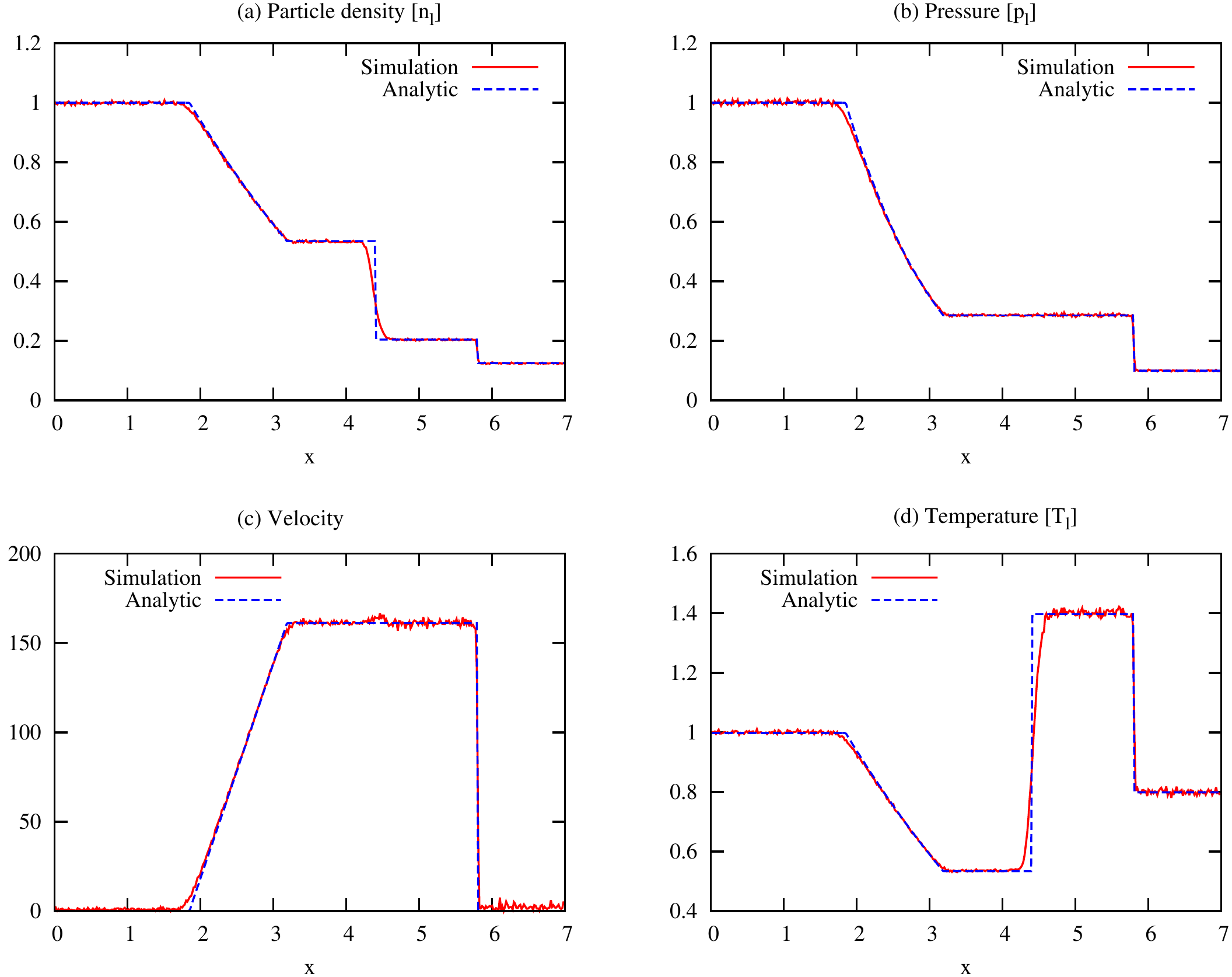}
\caption{2D Sod test as in Fig.~\ref{sod_2D}. Profiles of (a) normalized particle density, (b) normalized pressure, (c) bulk velocity, and (d) normalized temperature together with the analytic solutions.}
\label{sod_2D_profile}
\end{figure}
%%%%%%%%%%%%%%%%%%%%%%%
%%%%%%%%%%%%%%%%%%%%%%%
Fig.~\ref{sod_2D} shows the density, pressure, bulk velocity, and temperature per $o$-bin at a simulation time of $t \sim 5.6 \times 10^{-3}$ (time step 2200). To determine the profiles of $n$, $p$, $v_b$, and $T$ as functions of distance $x$, we average the corresponding properties over the y-dimension. The results are plotted in Fig.~\ref{sod_2D_profile} together with the analytic solutions. Density, pressure, and temperature are normalized by their values in the left side of the simulation space, $n_l \sim 2.906 \times 10^{6}$, $p_l \sim 1.305 \times 10^{11}$, and $T_l = 4.5 \times 10^4$, respectively. With 8 CPUs the 2D Sod test ran for a wall-time of about 21.6 hours. We find that the simulation agrees very well with the analytic solutions. Disagreement is present in the finite width of the contact discontinuity as can be seen in Figs.~\ref{sod_2D_profile}(a) and (d), which is also accompanied by a small increase in the bulk velocity. The latter has been observed by other kinetic approaches and was linked to the particle mean free path \cite{Crouseilles04,Chen11,Bennoune08}. Fluctuations are present in all quantities but are expected to decrease for larger $N$.  
%%%%%%%%%%%%%%%%%%%%%%%
%%%%%%%%%%%%%%%%%%%%%%%
\newline
For the 3D Sod test we keep the number of test-particles per $c$-bin the same as in the 2D simulation. However, a simple extension of the latter to 500 bins in the z dimension would result in $N = 10^{10}$. Such a high number of test-particles leads to a substantially larger memory requirement and, without a corresponding increase in the number of processors, a significantly longer computational time. Therefore, we set the number of $c$-bins to $400 \times 100 \times 100$ covering $0 \leq x \leq 7$ and $0 \leq y,z \leq 1.75$. With that, the number of test-particles becomes $N=8.0 \times 10^7$. For the output we apply $400 \times 100 \times 100$ $o$-bins while the particle mean free path is $\lambda = 10^{-3} \: \Delta x$, with $\Delta x = 7/400 = 1.75 \times 10^{-2}$.  
%%%%%%%%%%%%%%%%%%%%%%%
\begin{figure}
\centering
\includegraphics[width=0.9\textwidth]{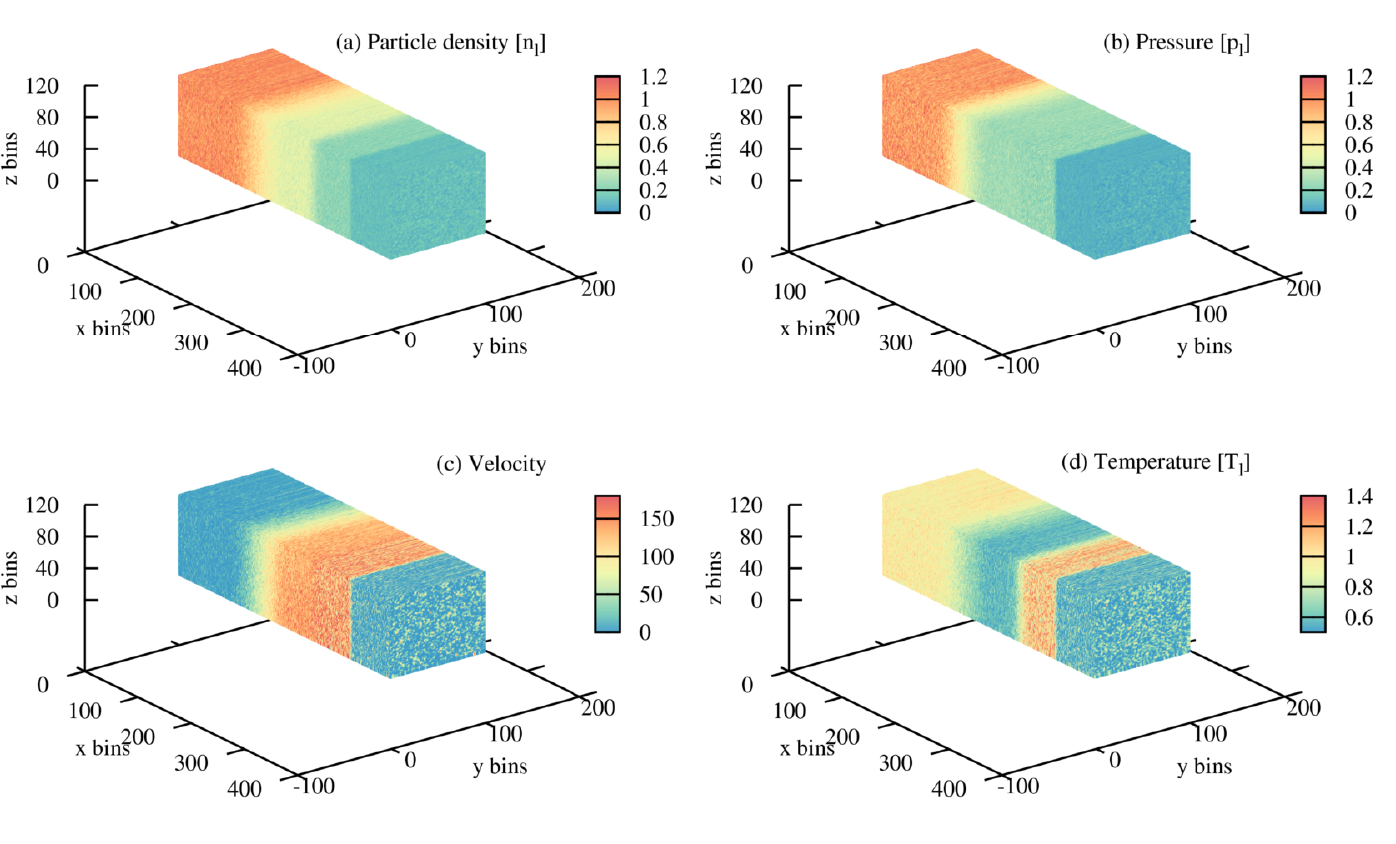}
\caption{3D Sod test: $N = 8.0 \times 10^7$ test-particles distributed over $400\times 100 \times 100$ bins with $\lambda = 10^{-3} \: \Delta x$. (a) Normalized particle density, (b) normalized pressure, (c) bulk velocity, and (d) normalized temperature per o-bin at time $t \sim 6.9 \times 10^{-3}$. Normalizations are done with $n_l \sim 6.6 \times 10^{6} $, $p_l \sim 1.99 \times 10^{12}$, and $T_l = 3 \times 10^{4} $.}
\label{sod_3D}
\vspace{1cm}
\centering
\includegraphics[width=0.8\textwidth]{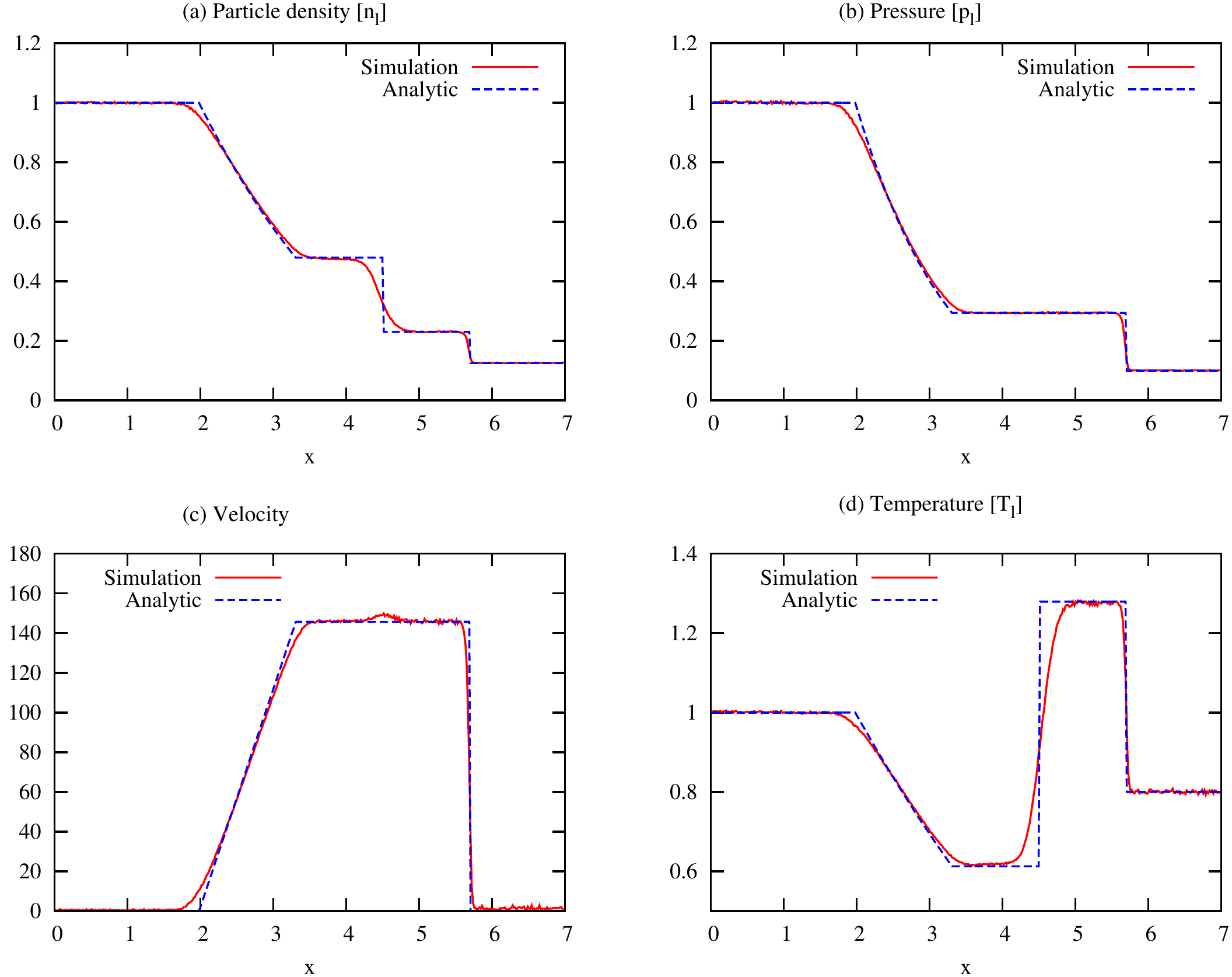}
\caption{3D Sod test as in Fig.~\ref{sod_3D}. Profiles of (a) normalized particle density, (b) normalized pressure, (c) bulk velocity, and (d) normalized temperature together with the analytic solutions.}
\label{sod_3D_profile}
\end{figure}
%%%%%%%%%%%%%%%%%%%%%%%
The resulting number density, pressure, radial velocity, and temperature per $o$-bin for $t \sim 6.9 \times 10^{-3}$ (time step 460) are shown in Fig.~\ref{sod_3D}. The profiles of all quantities are determined by their averages over bins in the $y$ and $z$ dimensions for a given distance $x$. The shock profiles are shown in Fig.~\ref{sod_3D_profile} together with the analytic solutions whereas normalizations are performed with $n_l \sim 6.6 \times 10^{7} $, $p_l \sim 1.99 \times 10^{12}$, and $T_l = 3 \times 10^{4} $. The results are qualitatively similar to the 2D simulation whereas the values of thermodynamic properties of the shocked matter are different due to the heat capacity ratio of $\gamma = 5/3$ for $f=3$. The fluctuations are smaller than in the 2D simulation due to the larger value of $N$. Note that the lower number of $c$-bins and therefore larger timestep size is the likely cause of the larger width of the contact discontinuity, as can be seen in Fig.~\ref{sod_3D_profile}(a) and (d). However, in general, we find that also the 3D Sod simulation reproduces the analytical solutions well. The 3D Sod test was performed with 30 CPUs whereas the wall-time for the simulation was 19.3 hours. While the timescales of the wall-time are thereby comparable to the 2D simulation, the total CPUs time is clearly higher due to the larger 
number of test-particles and $c$-bins.
%%%%%%%%%%%%%%%%%%%%%%%
%%%%%%%%               %%%%%%%%%%%
%%%%%%%%   NOH   %%%%%%%%%%%
%%%%%%%%               %%%%%%%%%%%
%%%%%%%%%%%%%%%%%%%%%%%
\subsection{Noh test}
\label{noh_test}
The Noh or Newtonian wall shock problem is an important test for codes that aim to study spherically collapsing systems such as CCSNe \cite{Noh87}. The test is initialized in form of gas with homogeneous density $n_0$ and zero pressure that streams with a radial velocity $v_{r,0}$ towards the origin of the simulation space. As infalling matter starts to pile up, it converts kinetic into internal energy. A shock front emerges at the origin, its radial distance growing as:
\begin{eqnarray}
r_{\mathrm{shock}} (t) = 0.5 \left( \gamma - 1 \right) v_{r,0} \: t. 
\end{eqnarray}
While the density of infalling matter evolves as: 
\begin{eqnarray}
n(r \geq r_{\mathrm{shock}}) &=& n_0 \left( 1 + \frac{v_{r,0}}{r} \: t \right)^{d - 1},
\end{eqnarray}
the density of matter which is enclosed by the shock is given by:
\begin{eqnarray}
n_{\mathrm{shock}} = n( r < r_{\mathrm{shock}}) &=& n_0 \left(\frac{\gamma + 1}{\gamma - 1}\right)^d. 
\label{noh}
\end{eqnarray}
Hereby, $d$ gives the geometry of the system with $d=1$ for planar, $d=2$ for cylindrical, and $d=3$ for spherical setups. The Noh test is very popular in numerical hydrodynamics as it can reveal anomalous wall-heating which is seen for many hydrodynamic codes and is caused by artificial viscosity. This excessive heating occurs at the walls of the simulation domain, increasing the temperature of matter and thereby decreasing its density far below the predicted value \cite{Noh87,Marti99}. The Noh test has not been studied by many particle based approaches and is therefore of special interest for our study. 
%%%%%%%%%%%%%%%%%%%%%%%
\begin{figure}
\centering
\includegraphics[width=0.8\textwidth]{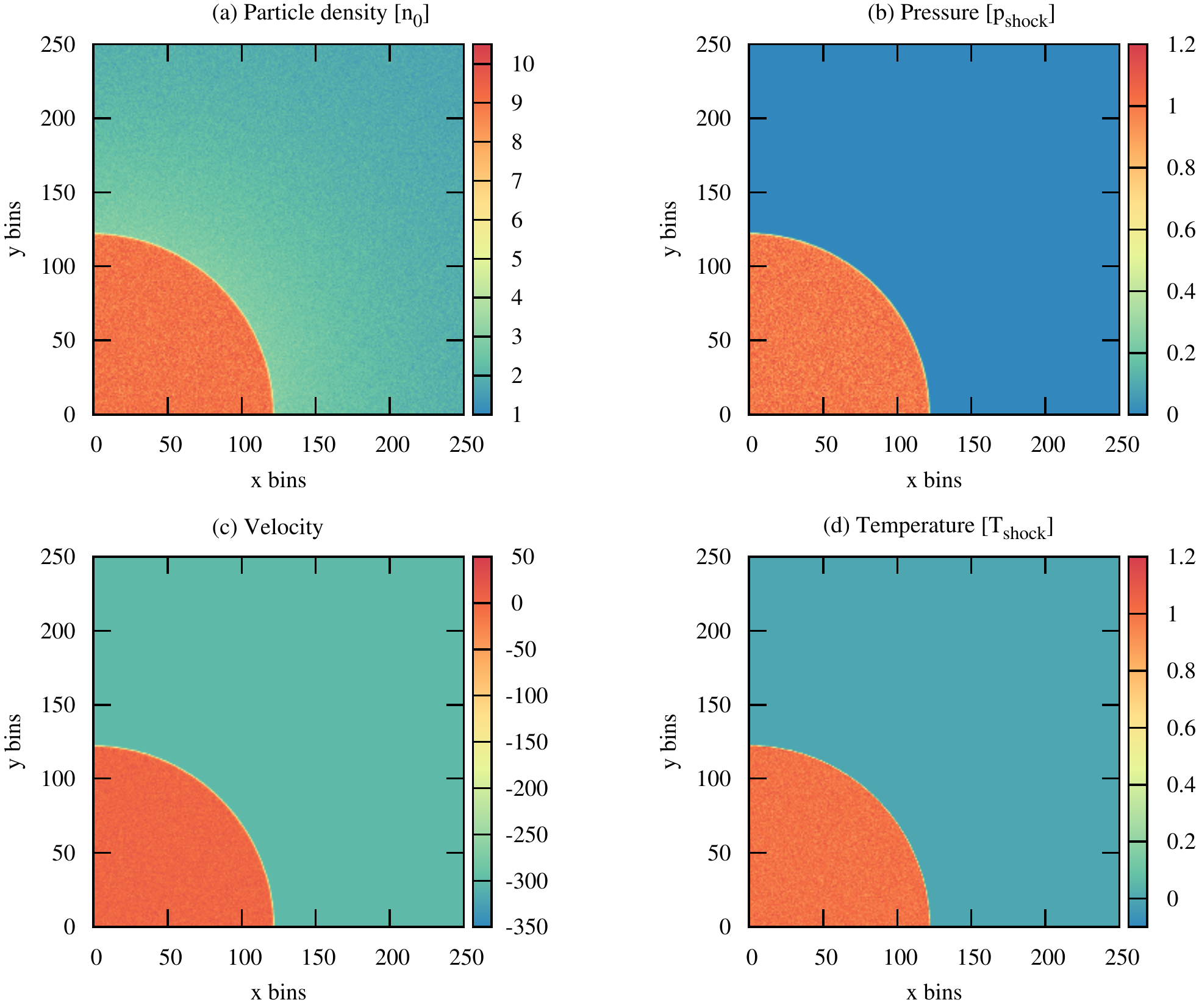}
\caption{Cylindrical Noh test in 2D: $N = 2.0 \times 10^7$ test-particles distributed over $500\times 500$ o-bins (only the first $250\times 250$ bins are shown).(a) Normalized particle density, (b) normalized pressure, (c) radial velocity, and (d) normalized temperature per o-bin at $t \sim 6.5 \times 10^{-3}$. Normalizations are done with $p_{\mathrm{shock}} \sim 5.06 \times 10^{11}$, $ T_{\mathrm{shock}} = 4.5 \times 10^4$, and $n_0 = 1.25 \times 10^6$.}
\label{noh_2D}
\vspace{1cm}
\centering
\includegraphics[width=0.8\textwidth]{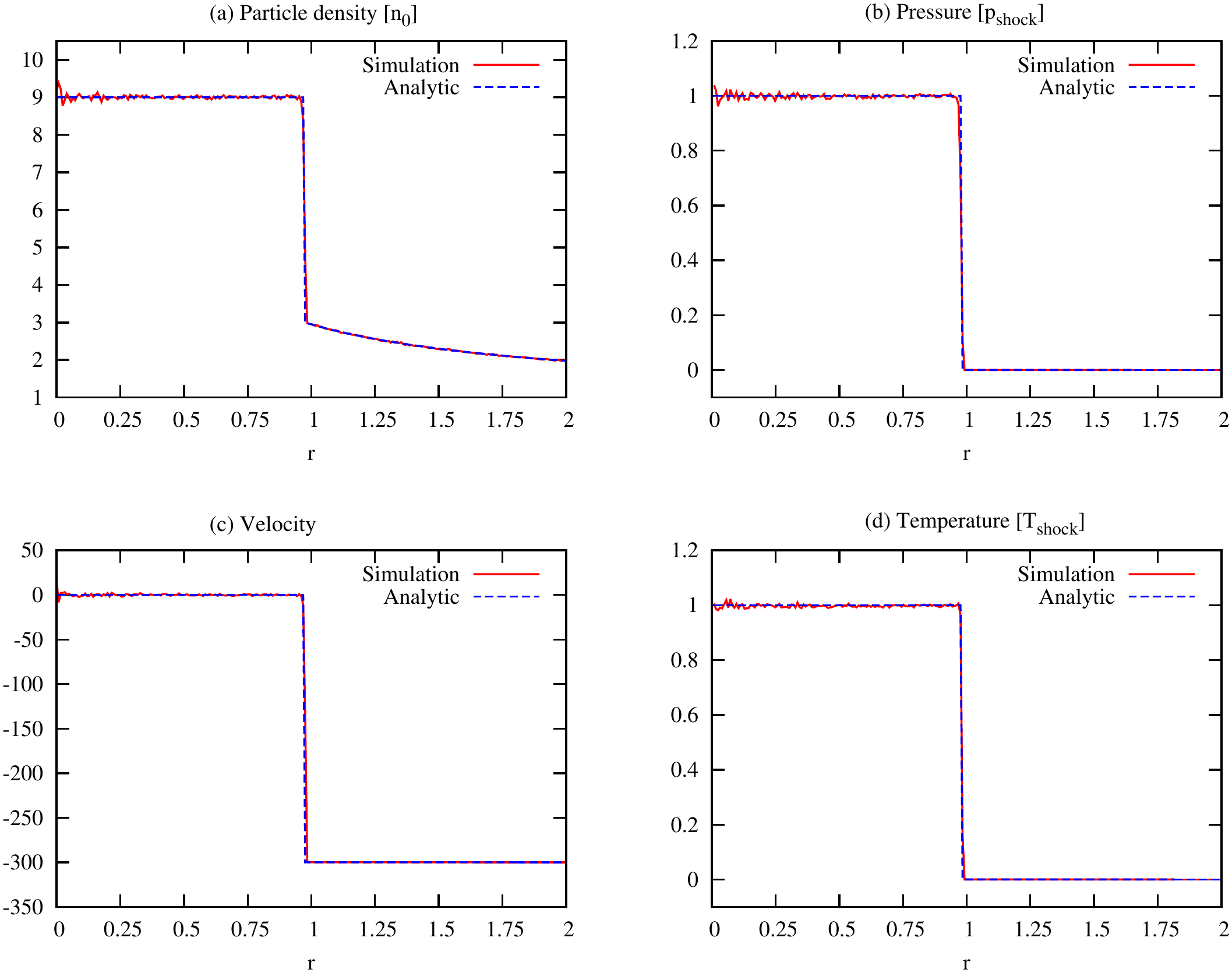}
\caption{Cylindrical Noh test in 2D as in Fig.~\ref{noh_2D}. Profiles of (a) normalized density, (b) normalized pressure, (c) radial velocity, and (d) normalized temperature as a function of distance $r$ together with the analytical solutions.}
\label{noh_2D_d2}
\end{figure}
%%%%%%%%%%%%%%%%%%%%%%%
%%%%%%%%%%%%%%%%%%%%%%%
\newline
We perform our tests in a 2D cylindrical ($d=2$) and 3D spherical ($d=3$) geometry. The results are compared to analytical solutions which are obtained by the noh.f code \cite{Fryxell_riemann}.  Figs.~\ref{noh_2D} and \ref{noh_2D_d2} show the results of the cylindrical Noh simulation with $2.0 \times 10^7$ test-particles at time $t \sim 6.5 \times 10^{-3}$ (time step 3700). The boundary conditions are again reflective and the simulation space is divided into $2000 \times 2000$ $c$-bins and $500 \times 500$ $o$-bins with $0.0 \leq x,y \leq 4.0$. As before, the particle mean free path is set to $\lambda = 10^{-3} \: \Delta x$ while $v_{r,0} = 300$. For the inner $250\times 250$ $o$-bins we plot the particle number density, pressure, radial velocity, and temperature per bin in Fig.~\ref{noh_2D}. The pressure and temperature are normalized by their values within the shock front $p_{\mathrm{shock}} \sim 5.06 \times 10^{11}$ and $T_{\mathrm{shock}} = 4.5 \times 10^4$ while the particle number density is divided by $n_0 = 1.25 \times 10^6$. The profiles of all quantities in Fig.~\ref{noh_2D_d2} are obtained as averages over the radial distance $r$. We find good agreement between numerical and analytical solutions. In the beginning of the simulation, when matter starts to pile up at the origin, we observe an initial overshoot and oscillations in the number density with $n_{\mathrm{shock}} \gtrsim 10\: n_0$. After several hundred time steps, the test-particles equilibrate and approach the predicted value of $n_{\mathrm{shock}} = 9 \: n_0$ (see eq.(\ref{noh})). Fluctuations are present in all quantities especially at the origin where particles are reflected on the walls. However, wall-heating, as it is present in many hydrodynamic codes, is never observed. Furthermore, the overall agreement between the analytical solution and the outcome of the simulation is remarkable.
%%%%%%%%%%%%%%%%%%%%%%%
\begin{figure}
\centering
\includegraphics[width=0.9\textwidth]{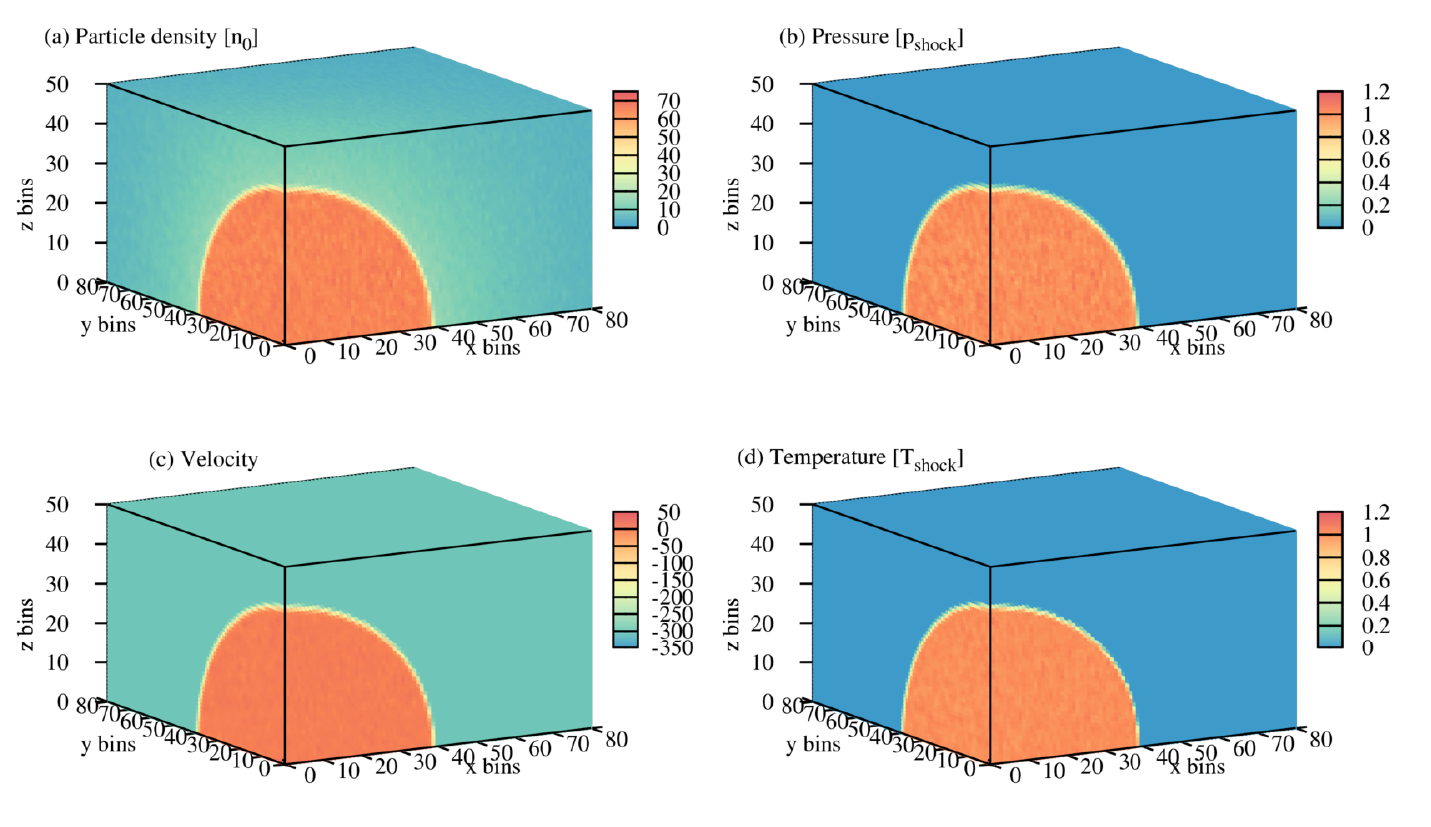}
\caption{Spherical Noh test in 3D: $N = 8.0 \times 10^7$ test-particles distributed over $200 \times 200 \times 200$ bins (only the first $80 \times 80 \times 50$ bins are shown). (a) Normalized particle density, (b) normalized pressure, (c) radial velocity, and (d) normalized temperature per bin at $t \sim 7.9 \times 10^{-3}$. Normalizations are done with $p_{\mathrm{shock}} \sim 2.4 \times 10^{12}$, $T_{\mathrm{shock}} = 3.0 \times 10^4$, and $n_0 = 1.25 \times 10^{6}$. }
\label{noh_3D}
\vspace{1cm}
\centering
\includegraphics[width=0.8\textwidth]{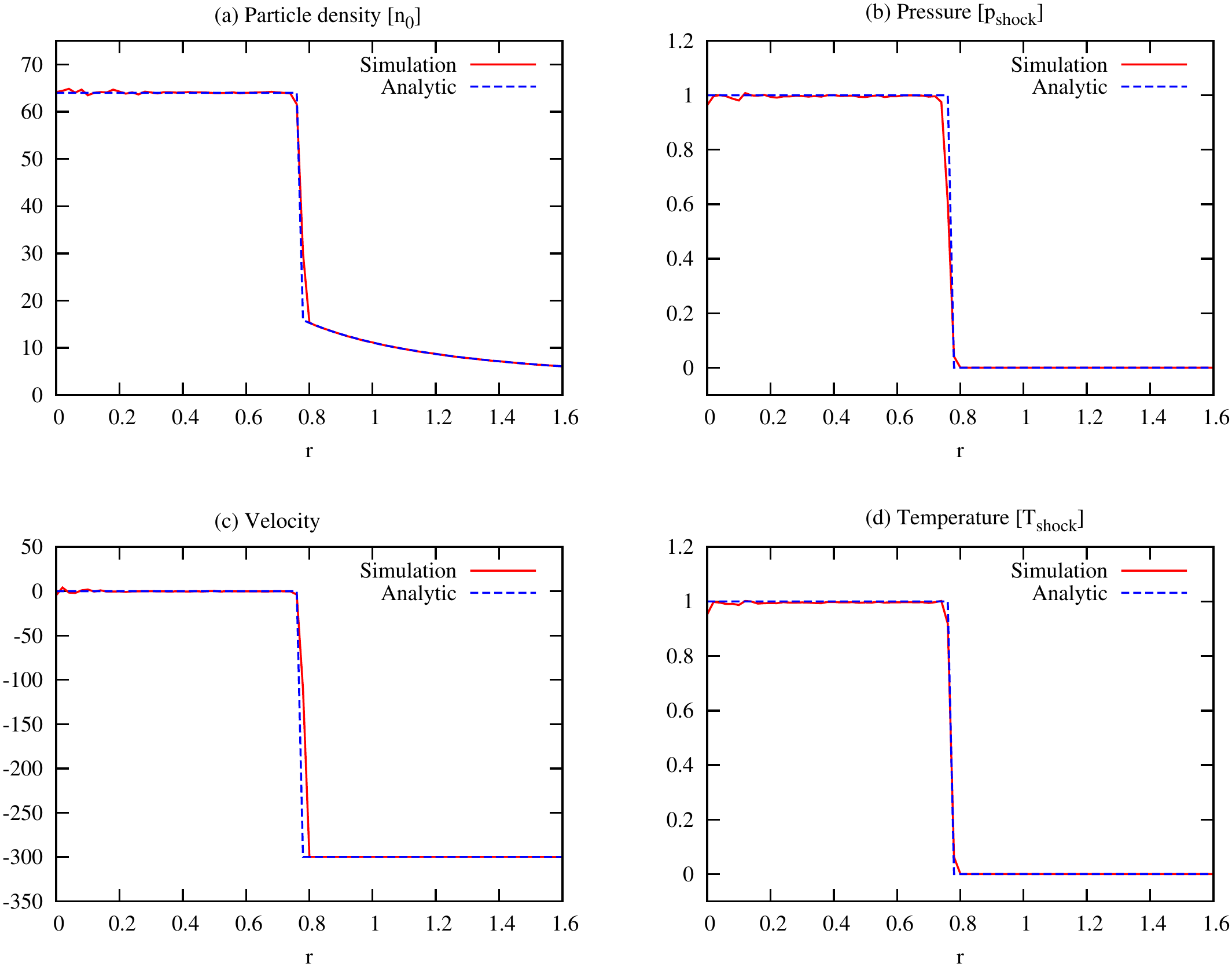}
\caption{Spherical Noh test in 3D as in Fig.\ref{noh_3D}: Profiles of (a) normalized particle density, (b) normalized pressure, (c) radial velocity, and (d) normalized temperature as a function of radial distance $r$ together with the analytical solutions.}
\label{noh_3D_profile}
\end{figure}
%%%%%%%%%%%%%%%%%%%%%%%
\newline
The spherical 3D Noh test is performed with $N = 8.0 \times 10^7$ test-particles in $400 \times 400 \times 400$ $c$-bins. For the output we use $200 \times 200 \times 200$ $o$-bins whereas the simulations space is again $0 \leq x,y,z \leq 4$ and $v_{r,0} = 300$. As before, the density at the origin overshoots the analytics solution at first but equilibrates within the first several hundred time steps. Fig.~\ref{noh_3D} shows the number density, pressure, radial velocity, and temperature at a simulation time $t \sim 7.9 \times 10^{-3}$ (time step 1500). We normalize the pressure, temperature, and number density by  $p_{\mathrm{shock}} \sim 2.4 \times 10^{12}$, $T_{\mathrm{shock}} = 3.0 \times 10^4$, and $n_0 = 1.25 \times 10^6$, respectively. The radial profiles of $n$, $p$, $v_r$ and $T$ are plotted in Fig.~\ref{noh_3D_profile}. The resolution of the shock front is lower than in the 2D simulation due to the lower number of $o$- and $c$-bins. The latter leads to a broadening of the shock front. However, the agreement between the analytical solution and the simulation is still good. Furthermore, again, no wall-heating is observed. While both, the 2D and 3D Noh tests evolve quickly in the beginning, they gradually slow down as particles pile up at the origin and their average number per $c$-bin increases from ca. 5 to 45 and from ca. 1.25 to 80 for the 2D and 3D tests, respectively. With 30 CPUs the 2D Noh simulation ran for a wall-time of about 17.8 hours while the duration of the 3D test was 34.8 hours. It should be noted that the large simulation time is mainly caused by the increase in particle numbers per $c$-bin. Collapse simulations like the Noh problem would greatly benefit from the implementation of an adaptive grid which could keep the number of particles per bin small.
%%%%%%%%%%%%%%%%%%%%%%%
%%%%%%%%               %%%%%%%%%%%
%%%%%%%% SEDOV%%%%%%%%%%%
%%%%%%%%               %%%%%%%%%%%
%%%%%%%%%%%%%%%%%%%%%%%
\subsection{Sedov-Taylor test}
\label{sedov_test}
Finally, we perform the Sedov-Taylor blast wave test. The initial setup of this problem is matter at uniform density, vanishing pressure and velocity. An explosion is set up via a point-like energy deposition $E_{\mathrm{blast}}$ in the center of the simulation space \cite{Taylor50,Sedov59}. The resulting spherically symmetric shock wave moves outwards with a radial distance given by:
\begin{eqnarray}
r_{\mathrm{shock}} = \left( \frac{E_{\mathrm{blast}} }{\alpha \: n_0 } \: t^2 \right)^{1/(2+d)} ,
\label{eq:sedov_r_shock}
\end{eqnarray}
whereas $\alpha$ is a constant of the order one with its exact value given by the heat capacity ratio $\gamma$. The latter also determines the number density of the shock front:
\begin{eqnarray}
n( r_{\mathrm{shock}}) = n_0 \: \left( \frac{\gamma + 1}{\gamma - 1} \right). 
\end{eqnarray}
The velocity of shocked matter has a radial dependence approximately $v_r \propto r/t$. As the shock wave moves away from the center it leaves behind matter at vanishingly low densities. With the pressure staying finite for $r=0$, the temperature grows and becomes infinitely large at the origin of the blast wave.    
%%%%%%%%%%%%%%%%%%%%%%%
%%%%%%%%%%%%%%%%%%%%%%%
\newline
Like the Noh test, the Sedov-Taylor problem is regarded as a standard test for hydrodynamic simulation codes, yet, it has not been performed with kinetic approaches. One challenging aspect is the initialization of the test. Neither hydrodynamic nor particle based simulations can operate with point-like energy depositions and have therefore to choose a finite size injection region. For hydrodynamic codes it has been shown that the size and shape of this region can impact the density, velocity, and pressure evolution of the shock front \cite{Fryer06,Sigalotti06}. The vanishingly small number densities at the center of the simulation space can be a challenge for codes operating with a finite number of particles as densities cannot become arbitrarily small \cite{Rosswog07}. Furthermore, some codes experience energy conservation problems during the first time steps of the simulation. This can cause the shock front to either run ahead or stay behind the analytic solution \cite{Tasker08}. Finally, a finite width of the shock front impacts the density profile and leads to a significant drop in the peak density. 
%%%%%%%%%%%%%%%%%%%%%%%
%%%%%%%%%%%%%%%%%%%%%%%
\newline
We perform the Sedov test in 2D and 3D with $d=2$ and $d=3$, respectively. In both cases we choose absorptive boundary conditions, i.e. particles do not interact with the walls of the simulation space. Once particles are beyond the spatial domain they are no longer evolved in the simulation. For the 2D test we choose again $N = 2.0\times 10^7$ test-particles distributed randomly in $2000\times 2000$ $c$-bins for $0 \leq x,y \leq 4.0$. To reduce fluctuations of thermodynamic properties in the central low density bins, we ensure a large initial number of test-particle per output bin and apply $250 \times 250$ $o$-bins. The mean free path is again $\lambda = 10^{-3} \: \Delta x$. 
%%%%%%%%%%%%%%%%%%%%%%%
\begin{figure}
\centering
\includegraphics[width=0.75\textwidth]{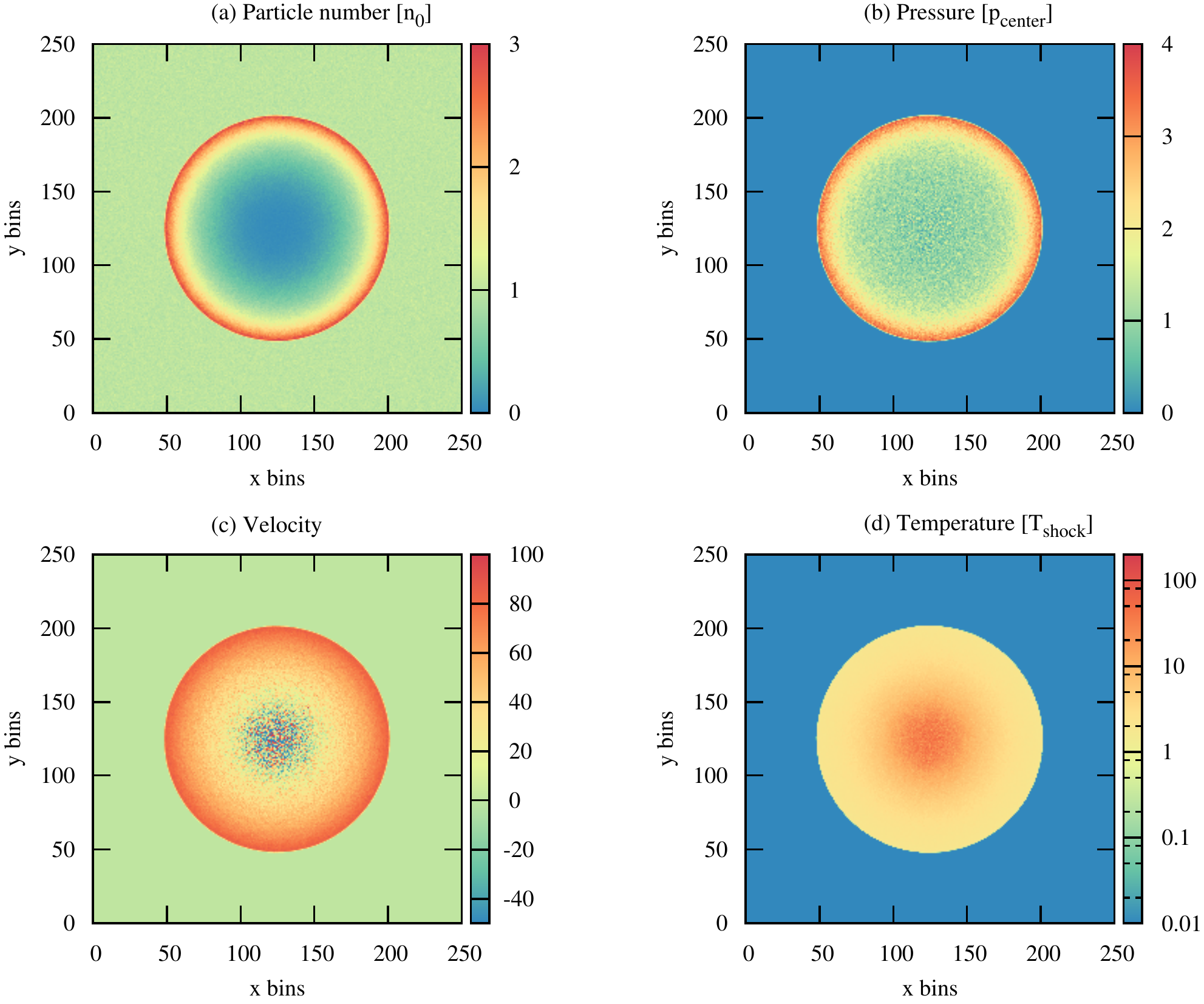}
\caption{Sedov test in 2D: $N = 2.0 \times 10^7$ test-particles distributed over $250\times 250$ $o$-bins with $\lambda = 10^{-3} \: \Delta x$. (a) Normalized particle density, (b) normalized pressure, (c) radial velocity, and (d) normalized temperature per bin at $t \sim 4.8 \times 10^{-3}$ }
\label{sedov_2D}
\vspace{1cm}
\centering
\includegraphics[width=0.8\textwidth]{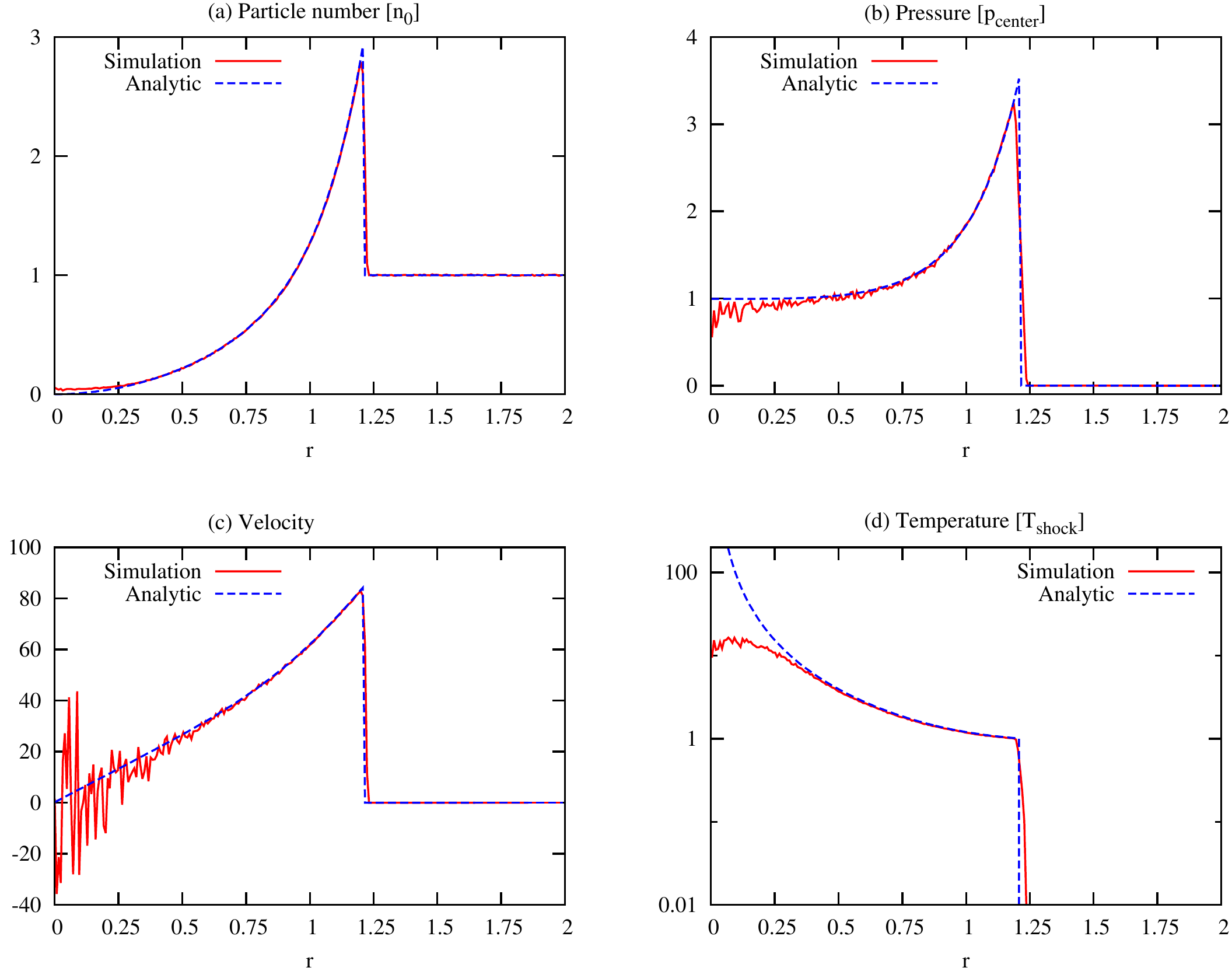}
\caption{Sedov test in 2D as in Fig.~\ref{sedov_2D}. Profiles of (a) normalized density, (b) normalized pressure, (c) radial velocity, and (d) normalized temperature together with the analytic solutions.}
\label{sedov_2D_profile}
\end{figure}
%%%%%%%%%%%%%%%%%%%%%%%
%%%%%%%%%%%%%%%%%%%%%%%
All particle velocity vectors are oriented randomly in space. The blast wave is set up in a region around $x=y=2.0$ with a radius of $r_{\mathrm{blast}} = 2 \: \Delta x$. Outside the blast region the absolute particle velocities are initialized as $v_{\mathrm{0}} = 10^{-3}$ while we deposit $E_{\mathrm{blast}}$ by applying $v_{\mathrm{blast}} =  40000$ for particles within $r \leq r_{\mathrm{blast}}$. With that, the resulting explosion energy is $E_{\mathrm{blast}} \sim 4.24 \times 10^{10}$. 
%%%%%%%%%%%%%%%%%%%%%%%
%%%%%%%%%%%%%%%%%%%%%%%
\newline
Within the first time steps after energy deposition, we see a density increase at the edges of the injection region. A spherical shock front forms and starts to propagate to larger $r$. Its peak density grows steadily while number densities at the center decrease to $n < 0.05 \: n_{0}$. Fig. \ref{sedov_2D} shows the number density, pressure, radial velocity, and temperature per bin at a simulation time of $t \sim 4.8 \times 10^{-3}$ (time step 4000). The particle number density is normalized by $n_0 = 1.25 \times 10^6$, while the pressure is divided by it's value at the center of the simulation space $p_{\mathrm{center}} \sim 3.75 \times 10^9$. The temperature is normalized by $T_{\mathrm{shock}} = 3.6 \times 10^3$ which is the value of $T$ at the shock front. It can be seen that the outgoing shock front is spherically symmetric. Fluctuations are mostly prominent in the pressure and the radial velocity. Both can be attributed to the low number of particles in these bins. Radial averages of $n$, $p$, $v_r$, and $T$ are shown in Fig.~\ref{sedov_2D_profile}, whereas the fluctuations in the pressure and radial velocity are again clearly visible. While $n$ reaches low values for $r \leq 0.25$ it cannot exactly reproduce the vanishingly small densities of the analytic solution. As can be seen in Fig.~\ref{sedov_2D_profile}(d) this has a direct impact on the temperature. The analytic solutions are well reproduced at large radial distances where $n$ is high, however, a clear under-prediction is present for the central region of the simulation space. Nevertheless, we find that the position of the shock front, its width, and the thermodynamic properties around the density peak match very well the analytic calculation. With 8 CPUs, the wall-time of the Sedov test is about 11.2 hours. This relatively short run-time in comparison to the previous tests is due to the small particle number per $c$-bin. The latter has a maximum value of ca. 15 particles per bin at the peak of the blast wave but stays otherwise at values $\lesssim 5$. 
%%%%%%%%%%%%%%%%%%%%%%%
\begin{figure}
\centering
\includegraphics[width=0.9\textwidth]{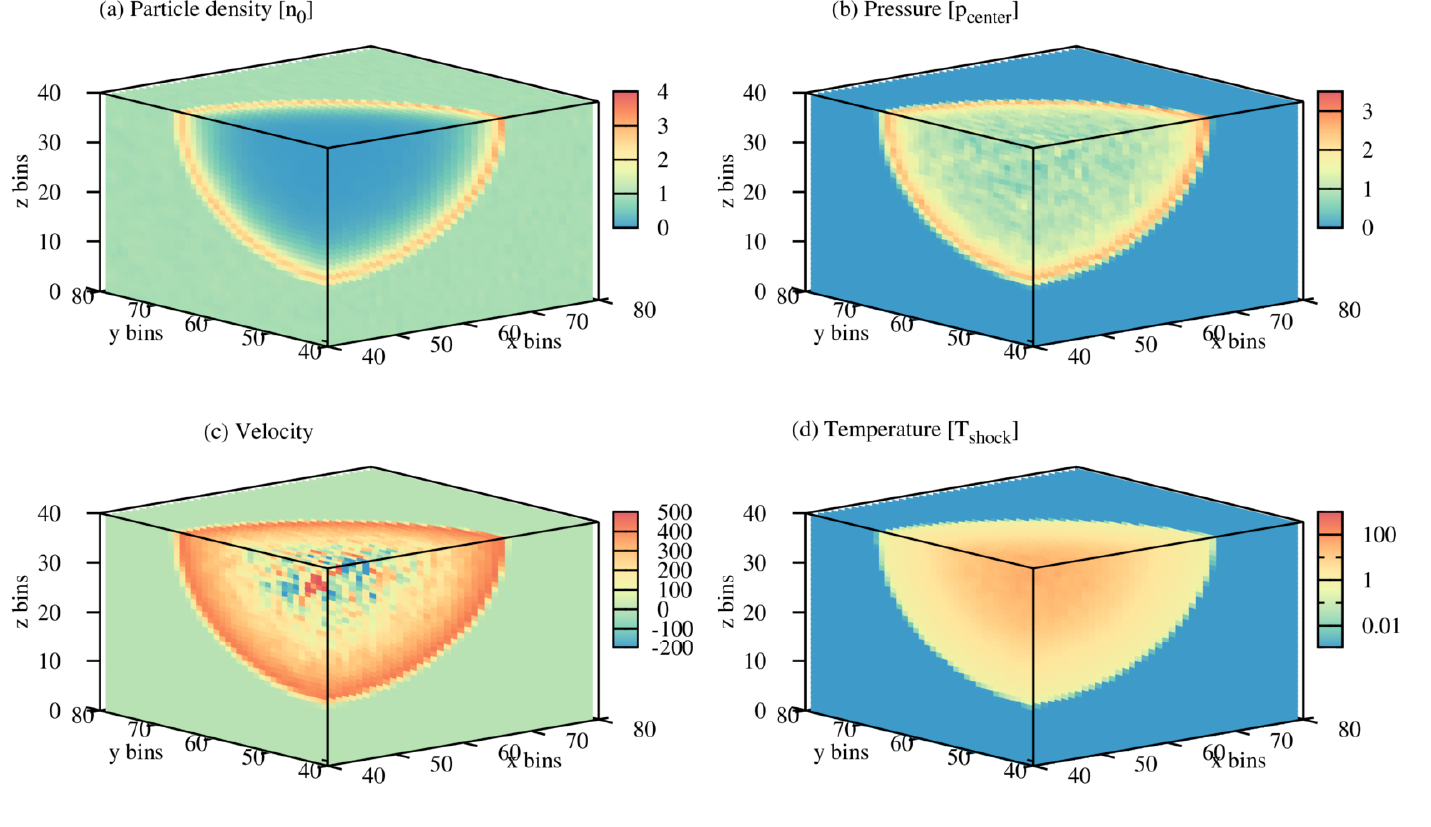}
\caption{Sedov test in 3D: $N = 2.0 \times 10^8$ test-particles distributed over $80\times 80 \times 80$ o-bins with $\lambda = 10^{-3} \: \Delta x$. (a) Normalized particle density, (b) normalized pressure, (c) radial velocity, and (d) normalized temperature per bin at $t \sim 9.1 \times 10^{-4}$. For better visualization, only one octant of the simulation space is shown. Normalizations are performed with $n_0 = 3.125 \times 10^6$, $p_{\mathrm{center}} \sim 2.5 \times 10^{11}$, and $T_{\mathrm{shock}} = 6.0 \times 10^4$.}
\label{sedov_3D}
\vspace{1cm}
\centering
\includegraphics[width=0.8\textwidth]{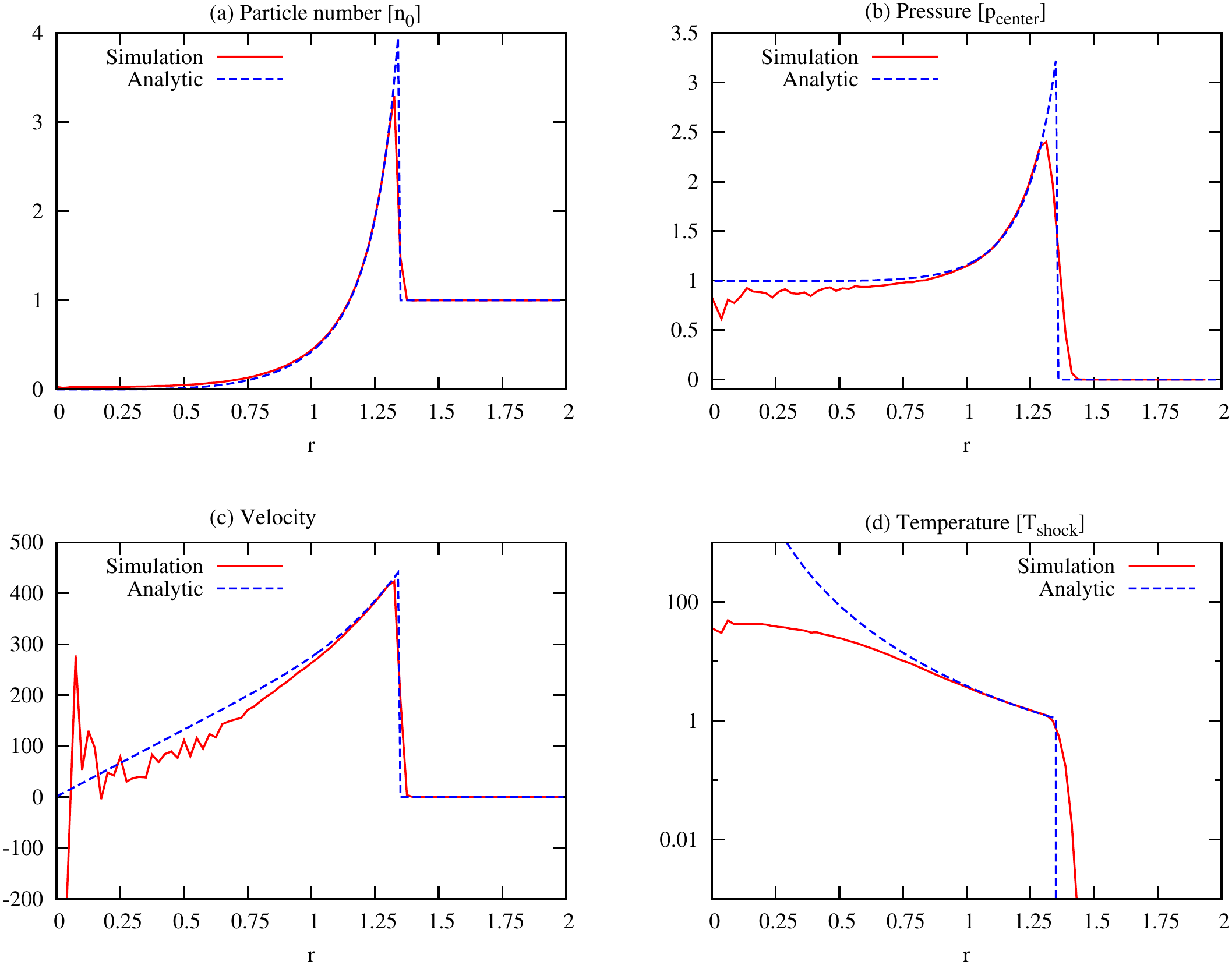}
\caption{Sedov test in 3D as in Fig.~\ref{sedov_3D}. Profiles of (a) normalized density, (b) normalized pressure, (c) radial velocity, and (d) normalized temperature together with the analytic solutions.}
\label{sedov_3D_profile}
\end{figure}
%%%%%%%%%%%%%%%%%%%%%%%
\newline
To ensure a large number of test-particles per $o$-bins in the 3D simulation as well as an acceptable number of $c$-bins for good resolution we choose $N = 2.0 \times 10^8$ with $400 \times 400 \times 400$ $c$-bins and $80 \times 80 \times 80$ $o$-bins. As in the 2D test, the size of the simulation space is $0 \leq x,y,z \leq 4$ whereas the blast wave is initialized at $x = y = z = 2$ with $r_{\mathrm{blast}} = 2 \: \Delta x$. Due to the larger number of particles in the injection region the blast energy is higher than in the 2D simulation with $E_{\mathrm{blast}} = 8.12 \times 10^{12}$. The results of the Sedov test are plotted in Figs.~\ref{sedov_3D} and \ref{sedov_3D_profile}. For better visualization, Fig.~\ref{sedov_3D}  shows the number density, pressure, radial velocity, and temperature per bin for only one octant of the simulation space. The normalizations are performed with $n_0 = 3.125 \times 10^6$, $p_{\mathrm{center}} \sim 2.5 \times 10^{11}$, and $T_{\mathrm{shock}} = 6.0 \times 10^4$. While fluctuations are again found to be prominent in the pressure and radial velocity, the number density and temperature per $o$-bin have a smooth behavior. However, Fig.~\ref{sedov_3D_profile} reveals that the radial average of $n$ again cannot match the low values of the analytic solution, especially for $r \lesssim 0.75$. A clear under-prediction of $p$, $v$, and $T$ can be found in Figs.~\ref{sedov_3D_profile}(b)-(d), whereas all deviations are more pronounced than in the 2D simulation, being most likely caused by the small number of test-particles per $c$-bin. As in the previous 2D simulations, the smaller number of bins leads to a larger width of the shock front and therefore decreased peak number density. However, despite these differences between the numerical and analytic results at low $n$, the position of the shock front and its thermodynamic properties are still in reasonable agreement with the predictions. Although the injection region is not strictly spherical, spherical symmetry of the shock front is preserved after initialization. Furthermore, as for all previous tests, the total energy in the simulation is determined at each time step and is found to be conserved. 
%%%%%%%%%%%%%%%%%%%%%%%
\newline
An increase in $N$ as well as the number of $o$-bins will most likely result in a smaller width of the shock front and a reduction of fluctuations, as well as the improvement regarding the under-prediction of $v_r$, $p$, and $T$. Nevertheless, generally, all particle approaches that are applied to problems which include a strong rarefaction of matter like the Sedov test will always be subject to the finite number of test-particles. However, as our work is focused on the simulation of core-collapse supernovae and ICF capsules where matter experiences a compression rather than a rarefaction our studies should not be largely affected by the performance of particle codes at $n \sim 0$. The run-time of the 3D Sedov test is higher than for the 2D problem which is mostly due to the larger value of $N$ and therefore total number of $c$-bins. With 30 CPUs the wall-time of the 3D test is 52.5 hours. As the particle number per $c$-bin does not increase largely during the simulation the run time of this test could be largely reduced via the application of more CPUs which can be achieved by enabling the use of distributed memory parallelization.
%%%%%%%%%%%%%%%%%%%%%%% 
%%%%%%%%%%%%%%%%%%%%%%%
%%%%%%%                  %%%%%%%%%%%
%%%%%%% NON-EQ %%%%%%%%%%%
%%%%%%%                  %%%%%%%%%%%
%%%%%%%%%%%%%%%%%%%%%%%
\section{Study of non-equilibrium systems}
\label{section:non_eq}
As mentioned in the introduction, an important property of kinetic schemes is their ability to evolve matter with large Knudsen numbers $K \gg 1$ that is not in hydrodynamic equilibrium. This property is of special interest for e.g. the study of shock waves and matter that contains particles with large mean free paths. In this section, we want to explore the performance of our code in this regime using values for $\lambda$ which are much larger than the characteristic length scale of the simulation space $L= \Delta x$. Accordingly, we chose representative particle mean free paths $\lambda = 1 \: \Delta x$, $10 \: \Delta x$, $30 \: \Delta x$, and $100 \: \Delta x$ and apply them to the previously discussed 2D Sod, Noh, and Sedov tests, comparing the results of the simulation to the analytic solution as well as the free-streaming regime. In the latter, particles can only interact with the walls of the simulation space but not with each other and $\lambda$ can therefore be regarded as infinitely large. While mean free path studies have been performed for the Sod test \cite{Bennoune08,Bouras10}, where results are compared to viscous hydrodynamic simulations, we are not aware of similar work that applied either the Noh or the Sedov problem. The setup and initialization of all tests is the same as in the continuum regime described in sections \ref{sod_test}, \ref{noh_test}, and \ref{sedov_test}, whereas for the output we only analyze the spatial profiles of $n$, $v$, $p$, and $T$. 
%%%%%%%%%%%%%%%%%%%%%%%
\newline
The 2D Sod test is performed with $N=2.0 \times 10^7$ test-particles with $0 \leq x\leq 7$ and $0 \leq y\leq 1.75$, $2000 \times 500$ $c$-bins, and $400 \times 100$ $o$-bins.
%%%%%%%%%%%%%%%%%%%%%%%
\begin{figure}
\centering
\includegraphics[width=0.8\textwidth]{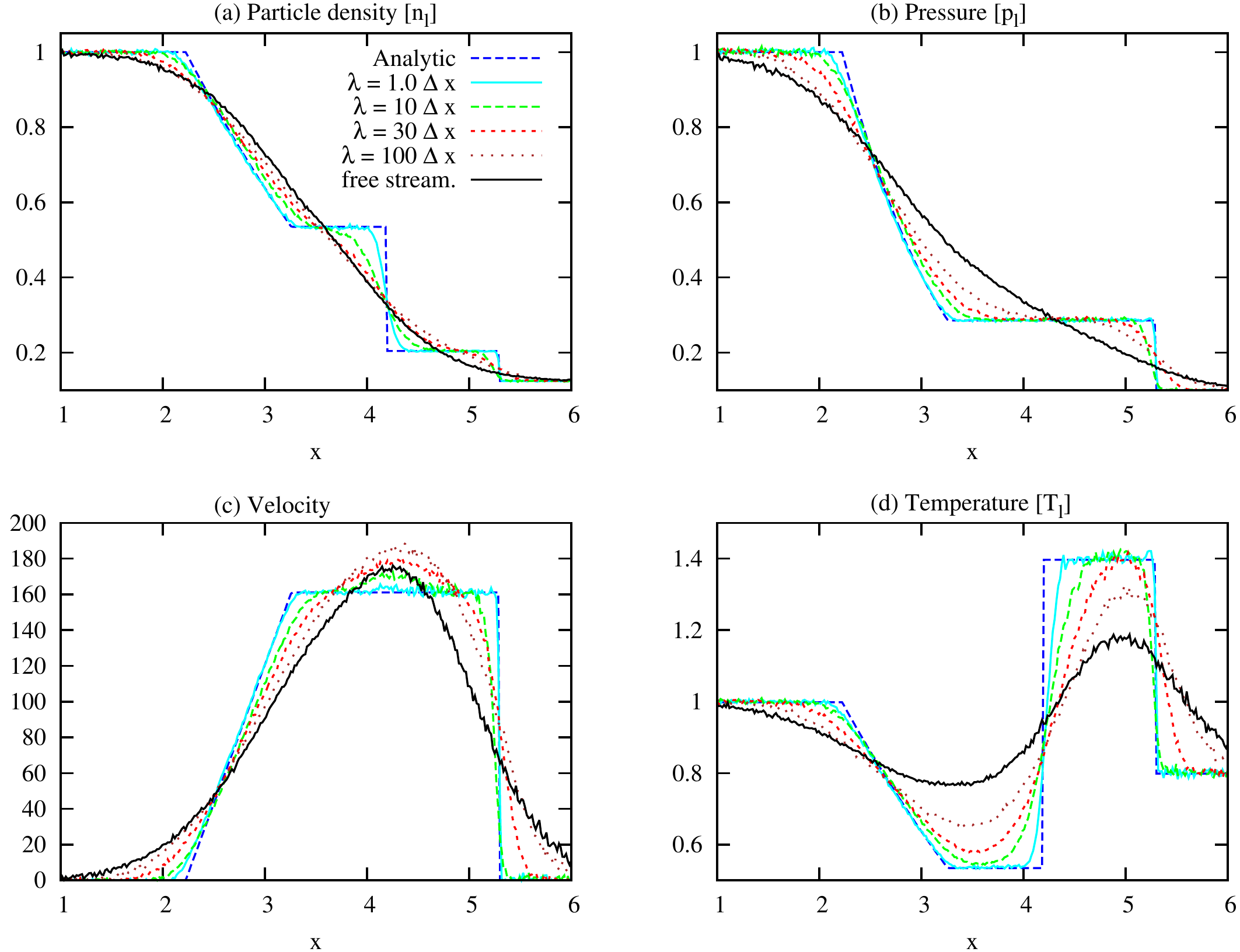}
\caption{2D Sod mean free path study. Profiles of (a) normalized number density, (b) normalized pressure, (c) bulk velocity, and (d) normalized temperature are plotted together with the analytic solution at a simulation time $t \sim 4.3 \times 10^{-3}$ for $50 \leq x \leq 350$.The mean free path is varied from $\lambda=1.0 \: \Delta x$, $10 \: \Delta x$ to $30 \: \Delta x$, to $100 \Delta x$. In addition, a simulation using non-interacting particles is shown.}
\label{sod_2D_mfp}
\end{figure}
%%%%%%%%%%%%%%%%%%%%%%%
Fig.~\ref{sod_2D_mfp} shows the results with different values of $\lambda$ at a simulation time of $t \sim 4.3  \times 10^{-3}$. As has been discussed in Sect.~\ref{maxwell_boltzmann}, despite a lower interaction rate, a mean free path of $\lambda = 1 \: \Delta x$ can reproduce the continuum solution well. Since the collision neighborhood has a size of $3 \: \Delta x \times 3 \: \Delta x$, many particles with $\lambda = 1 \: \Delta x$ succeed in finding collision partners and thereby ensure  a prompt hydrodynamic equilibration. As it can be seen in Fig.~\ref{sod_2D_mfp}, with larger mean free paths the shock front and the contact discontinuity become progressively washed out. While the general features of the continuum pressure profile are reproduced up to $\lambda = 30 \: \Delta x$, the density seems to be stronger affected by the increasing value of $\lambda$ and is more similar to the free streaming regime.  
%%%%%%%%%%%%%%%%%%%%%%%
%%%%%%%%%%%%%%%%%%%%%%%
\begin{figure}
\centering
\includegraphics[width=0.8\textwidth]{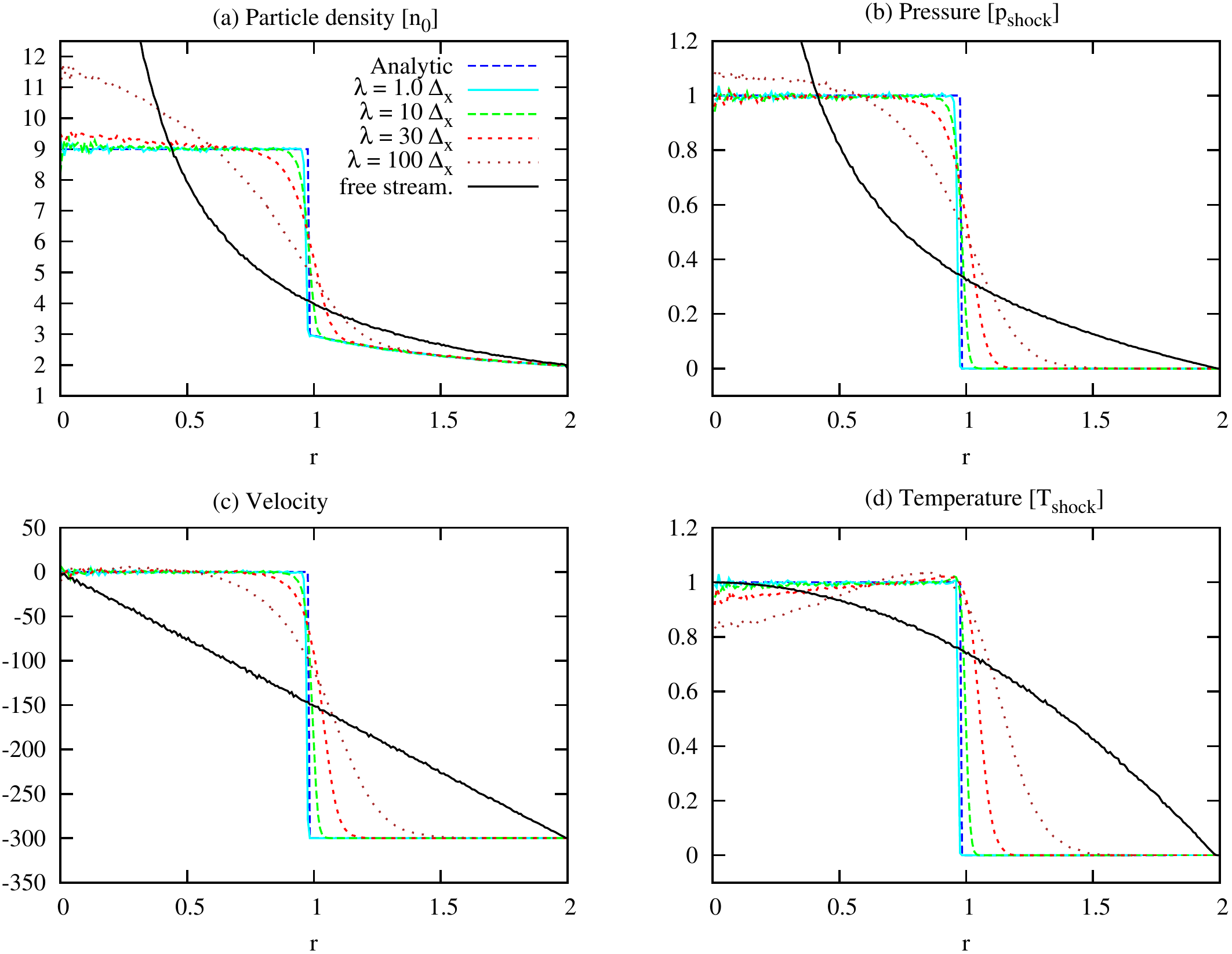}
\caption{2D Noh mean free path study. Profiles of (a) normalized number density, (b) normalized pressure, (c) radial velocity, and (d) normalized temperature are plotted together with the analytic solution at a simulation time $t \sim 6.5 \times 10^{-3}$.The mean free path is varied from $\lambda=1.0 \: \Delta x$, $10 \: \Delta x$ to $30 \: \Delta x$, to $100 \Delta x$. In addition, a simulation using non-interacting particles is shown.}
\label{noh_2D_mfp}
\end{figure}
%%%%%%%%%%%%%%%%%%%%%%%
%%%%%%%%%%%%%%%%%%%%%%%
\begin{figure}
\centering
\includegraphics[width=0.8\textwidth]{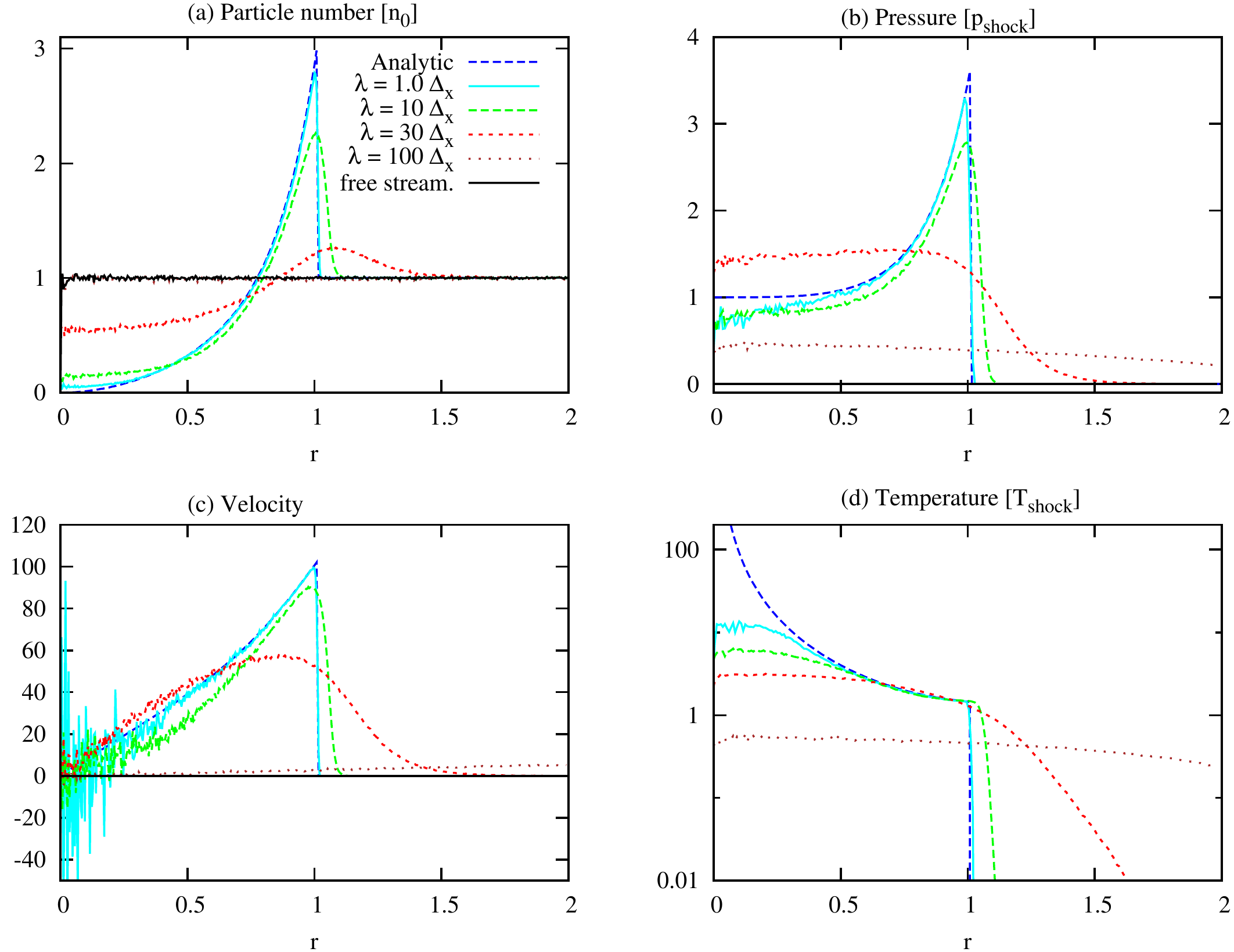}
\caption{2D Sedov mean free path study. Profiles of (a) normalized number density, (b) normalized pressure, (c) radial velocity, and (d) normalized temperature are plotted together with the analytic solution at a simulation time $t \sim 3.3 \times 10^{-3}$.The mean free path is varied from $\lambda=1.0 \: \Delta x$, $10 \: \Delta x$ to $30 \: \Delta x$, to $100 \Delta x$. In addition, a simulation using non-interacting particles is shown.}
\label{sedov_2D_mfp}
\end{figure}
%%%%%%%%%%%%%%%%%%%%%%%
\newline
The results of the Noh mean free path study are shown in Fig.~\ref{noh_2D_mfp}. For the output, we plot $n$, $p$, $v_r$, and $T$ as functions of radial distance $r$ whereas the simulation setup is the same as in section \ref{noh_test}. Like in the Sod test, the simulation using $\lambda = 1.0 \: \Delta x$ matches the analytic solution well, whereas with increasing values of $\lambda$, a clear broadening of the shock front can be seen. In the continuum limit, particles pile up at one corner of the simulation space and are trapped by in-flowing matter. They equilibrate and lead to the predicted continuum density, pressure, velocity, and temperature profiles. However, as $\lambda$ is increased, a growing number of particles is no longer trapped and can escape. Fewer incoming particles are stopped at the shock and can thereby travel further into the shocked matter region. Both effects result in a broadening of the shock front. With the number of incoming particles being larger than the number of particles moving outwards, $n_{\mathrm{shock}}$ steadily increases and rises above the analytic solution. This increase is more pronounced for higher values of $\lambda$ and reaches its extreme for free streaming matter. The smaller increase in the pressure in comparison to $n$ leads to the temperature dip which can be seen in Fig.~\ref{noh_2D_mfp}(d). With larger mean free paths, the particle densities and radial velocities show a clear approach towards the free streaming regime. In the latter, velocities of inflowing and outgoing particles add up and result in the observed radial dependence of $v_r$. For the temperature profiles, a clear difference is present between the free streaming regime and $\lambda = 100 \: \Delta x$. Since particles are initialized with the same radial velocities rather than the Maxwell-Boltzmann velocity distribution, matter in the free streaming regime never reaches equilibrium. With the lack of the latter, a temperature cannot be clearly defined and a comparison of $T$ between free streaming matter and matter with $\lambda = 100 \: \Delta x$ is therefore not possible.
%%%%%%%%%%%%%%%%%%%%%%%
\newline
The Sedov problem is another interesting test case for our mean free path study. Possibly stronger than in the Sod and the Noh tests, its shock wave formation relies on the frequent interaction of particles. Initially, only a few test-particles in the injection region receive the blast wave energy. In the continuum limit, frequent collisions distribute the latter among other particles which leads to the formation of a spherical shock front. As the shock propagates to larger radii it pushes matter outwards. The peak density of the shock is very sensitive to its width and can decrease significantly if the latter is not sufficiently small. A large particle mean free path can therefore directly affect the shock wave structure either via a simple broadening of the front, a delay of its formation to larger radial distances $r$, or no shock wave formation at all. If high energy particles from the injection region are free streaming and therefore prohibited from interaction, they simply propagate to larger radii and eventually leave the simulations space without depositing their energy in the surrounding medium. As a consequence, the overall particle density, pressure, velocity, and temperature remain almost unaffected.   
%%%%%%%%%%%%%%%%%%%%%%%
\newline
As for the previous tests, the setup of the Sedov test is the same as in the continuum regime in Sect.\ref{sedov_test}, with $N = 2.0 \times 10^7$, $250 \times 250$ $o$-bins, and $0 \leq x,y \leq 4$. The radial averages of $n$, $p$, $v_r$, and $T$ with different values of $\lambda$ are obtained at a simulation time of about $t \sim 3.3 \times 10^{-3}$ and plotted in Fig.~\ref{sedov_2D_mfp} together with the analytic solution and the free streaming scenario. Like in the Sod and the Noh test, we find that the simulation using $\lambda = 1.0 \: \Delta x$ is very close to the analytic prediction whereas deviations, for example in form of fluctuations at the origin, are very similar to the ones discussed in Sect. \ref{sedov_test}. For $\lambda = 10 \: \Delta x$ we see a broadening of the shock front accompanied by the expected decrease in the shock peak density. At the same time, densities at smaller radial distances $r$ are higher in comparison to the analytic solution and the simulation with $\lambda = 1.0 \: \Delta x$. With decreasing particle interaction rates, less particle are pushed outwards by the shock front and consequently stay behind at smaller $r$. For the free streaming regime and $\lambda = 100 \: \Delta x$ we find that the density distribution seems to remain unaltered in comparison to its state at $t=0$. However, for $\lambda = 100 \: \Delta x$ we find that the pressure, radial velocity, and temperature experience a small increase above their initial values. As already mentioned, the Sedov blast wave is strongly dependent on the particle interaction and energy distribution shortly after its initialization in the injection region. When applying large particle mean free paths, this energy distribution does not take place and the fast particles from the injection region quickly escape the simulation space. This is the scenario which we observe for the free streaming particles in Fig.~\ref{sedov_2D_mfp}. The lack of any features in the density, velocity, pressure, and temperature profiles shows that the particles from the injection region have already escaped the simulation space. For $\lambda = 100 \: \Delta x$ interactions take place but are not sufficient to initialize a bulk motion of matter in form of a shock front. However, as the high energy particles travel to larger radii they distribute some of their kinetic energy along the way via occasional particle scattering. The interaction frequency is not high enough to lead to an observable density enhancement, however, in comparison to the free streaming regime, the large particle velocities result in slightly higher values of $p$, $v_r$, and $T$. The simulation using $\lambda = 30 \: \Delta x$ shows a shock broadening, similar, but more pronounced than for $\lambda = 10 \: \Delta x$. In addition, fast particles that can escape to large $r$ result in an enhanced of the temperature almost up to the edge of the simulation space. The relatively large pressure at small $r$ in comparison to $\lambda = 1.0 \: \Delta x$ and $\lambda = 10 \: \Delta x$ is most likely a result of the correspondingly larger number of test-particles in this region, as can be seen in Figs.~\ref{sedov_test}(b) and (c). As for $\lambda = 100 \: \Delta x$, these particles have not been pushed out by the shock wave but stayed behind with larger velocities due to the rare interaction with the particles from the injection region. 
%%%%%%%%%%%%%%%%%%%%%%%
\newline
In summary we find that our mean free path study of the Sod, Noh, and Sedov tests reveal a shock broadening with larger values of $\lambda$ as has been seen in other kinetic codes and viscous hydrodynamic simulations of the Sod test. However, at the same time, clear non-equilibrium effects can be observed for $K \gg 0$, for example in the Sedov test, where fast particles from the energy injection region leave the simulation space only scarcely interacting with their surrounding and thereby highly altering the structure of $n$, $p$, $v$, and $T$. In general, the timing results of the performed non-equilibrium tests are similar to the results obtained in sections \ref{sod_test}, \ref{noh_test}, and \ref{sedov_test}. For increasing values of $\lambda$ we find again a decrease in computational time as it was discussed in the Maxwell-Boltzmann equilibration study in section \ref{maxwell_boltzmann}.
%%%%%%%%%%%%%%%%%%%%%%%%%%%%%%%%%%%%%%%%%%%%%%%%%%%%%%%%%%%%%%%%%%%%%%
%%%%%%%%%%%%%%%%%%%%%%%%%%%%%%%%%%%%%%%%%%%%%%%%%%%%%%%%%%%%%%%%%%%%%%
\section{Summary and Outlook}
\label{section:conclusion}
In this work we introduce and discuss a modified Monte Carlo kinetic scheme which simulates the phase space evolution of systems that can move in and out of hydrodynamic equilibrium. The code operates with test-particles that interact with each other via two-body collisions with variable particle mean free paths. To achieve high computational efficiency, the code is written in a parallel form and applies spatial binning of test-particles. We determine the interaction partners using the Point of Closest Approach technique with the final collision partners being chosen according to the shortest distance between them. Both deviations from the usual Direct Simulation Monte Carlo techniques, which determine interaction partners randomly from a scattering cell, are aimed to enhance spatial accuracy which could otherwise be compromised due to scattering partners which are far away from each other. For first performance studies of our kinetic scheme we apply timing and mean free path tests with a simple gas of test-particles moving in two dimensions and interacting with each other via binary collisions. With equal initial absolute values, the test-particle velocities quickly equilibrate to the Maxwell-Boltzmann distribution when the particle mean free path is chosen to be small with respect to the distance which can be travelled by a particle within one timestep. Larger mean free paths on the other hand lead to longer equilibration times. In addition, we find that the computational speed-up of the scattering partner search shows almost ideal behavior for both, simple 2D and 3D simulations. 
%%%%%%%%%%%%%%%%%%%%%%%
\newline
To further explore the capability of our kinetic scheme to reproduce hydrodynamic behavior as well as handle steep gradients and discontinuities, we apply a suite of shock wave tests consisting of the Sod, the Noh, and the Sedov problems. All simulations are performed in 2D and in 3D with tens of millions of test-particles. The evolution of the resulting shock waves is compared to the corresponding analytic solutions, whereas we find good to excellent agreement between the predicted shock profiles and the results of the kinetic simulations. Finally, we test the ability of the kinetic particle scheme to evolve systems which are not in hydrodynamic equilibrium. We perform all three shock tests applying particle mean free paths which are larger than the distance which a test-particle can travel within one timestep. As expected, we find that larger particle mean free paths lead to a more pronounced broadening of the shock front which is influenced by non-equilibrium effects. We can conclude that the particle kinetic scheme is capable to reproduce hydrodynamic behavior, especially with regard to shock wave evolution. More importantly, it can also simulate systems which are not in the continuum regime. With this, it is an interesting tool to study systems such as core-collapse supernovae where neutrinos traverse from the trapped to the free streaming regime as well as the dynamics of inertial confinement fusion capsules where components of the nuclear fuel, such as thermal deuterons, posses mean free paths which are larger than the characteristic length scales of the thermonuclear ignition region so that a hydrodynamic regime cannot be assumed.  
%%%%%%%%%%%%%%%%%%%%%%%
\newline
Future applications of our code in e.g. core-collapse supernova simulations will require further performance studies such as additional shock tests in two and three dimensions as well as studies concerning the ability of the particle approach to reproduce turbulent fluid motion, for example Rayleigh-Taylor and the Helmholtz instabilities. The search for scattering partners has to be upgraded to be able to handle particles moving at relativistic speeds. Also, realistic three dimensional simulations will require higher particle numbers than were applied in the current study of $N \sim 10^9$. To handle such a large amount of particles, the scattering partner search has to be set up with distributed memory parallelization to enable the usage of $\gg 10^2$ processors which is currently in development. Corresponding timing tests have to be performed whereas large simulations would benefit from additional extensions such as an adaptive grid to ensure a homogeneous computational load of processors. Furthermore, future applications  will implement realistic mean field forces and scattering cross-sections, for example nuclear forces and  neutrino interactions with nuclear matter in core-collapse supernovae, as well as the handling of long-range forces such as gravitation. Finally, we plan to provide the scattering partner search code in form of a library to be applicable in a variety of studies of different hydrodynamic and transport problems. 
%%%%%%%%%%%%%%%%%%%%%%%%%%%%%%%%%%%%%%%%%%%%%%%%%%%%%%%%%%%%%%%%%%%%%%
%%%%%%%%%%%%%%%%%%%%%%%%%%%%%%%%%%%%%%%%%%%%%%%%%%%%%%%%%%%%%%%%%%%%%%
\section{Acknowledgements}
The authors would like to thank the Blue Water Undergraduate Petascale Education Program and Shodor for their financial and educational support. Furthermore, this work used the Extreme Science and Engineering Discovery Environment (XSEDE), which is supported by National Science Foundation grant number OCI-1053575. I.S. is thankful to the Alexander von Humboldt foundation and acknowledges the support of the High Performance Computer Center and the Institute for Cyber-Enabled Research at Michigan State University. T.S. is grateful for useful conversations with LANL physicists James Cooley and James Mercer-Smith that helped guide his contribution to this work.

\bibliographystyle{ieeetr}
\bibliography{references}

\end{document}